\documentclass[a4paper,onecolumn,11pt,accepted=2022-12-10]{quantumarticle}
\pdfoutput=1
\usepackage[utf8]{inputenc}
\usepackage[english]{babel}
\usepackage[T1]{fontenc}
\usepackage{amsmath}
\usepackage{hyperref}

\usepackage{tikz}
\usepackage{lipsum}

\begin{document}

\title{Quantum Liouvillian exceptional and diabolical points
for bosonic fields with quadratic Hamiltonians: The
Heisenberg-Langevin equation approach}

\author{Jan Pe\v{r}ina Jr.}
\orcid{0000-0003-0542-7508} \email{jan.perina.jr@upol.cz}
\affiliation{Joint Laboratory of Optics of Palack\'{y} University
and Institute of Physics of CAS, Faculty of Science, Palack\'{y}
University, 17. listopadu 12, 771 46 Olomouc, Czech Republic}
\orcid{0000-0002-2445-2701}
\author{Adam Miranowicz}
\orcid{0000-0002-8222-9268}
\affiliation{Institute of Spintronics
and Quantum Information, Faculty of Physics, Adam Mickiewicz
University, 61-614 Pozna\'{n}, Poland}
\author{Grzegorz Chimczak}
\orcid{0000-0003-2981-4027}
\affiliation{Institute of Spintronics
and Quantum Information, Faculty of Physics, Adam Mickiewicz
University, 61-614 Pozna\'{n}, Poland}
\author{Anna Kowalewska-Kud\l{}aszyk}
\orcid{0000-0002-6088-099X}
\affiliation{Institute of Spintronics
and Quantum Information, Faculty of Physics, Adam Mickiewicz
University, 61-614 Pozna\'{n}, Poland}

\maketitle

\begin{abstract}
Equivalent approaches to determine eigenfrequencies of the
Liouvillians of open quantum systems are discussed using the
solution of the Heisenberg-Langevin equations and the
corresponding equations for operator moments. A simple damped
two-level atom is analyzed to demonstrate the equivalence of both
approaches. The suggested method is used to reveal the structure
as well as eigenfrequencies of the dynamics matrices of the
corresponding equations of motion and their degeneracies for
interacting bosonic modes described by general quadratic
Hamiltonians. Quantum Liouvillian exceptional and diabolical
points and their degeneracies are explicitly discussed for the
case of two modes. Quantum hybrid diabolical exceptional points
(inherited, genuine, and induced) and hidden exceptional points,
which are not recognized directly in amplitude spectra, are
observed. The presented approach via the Heisenberg-Langevin
equations paves the general way to a detailed analysis of quantum
exceptional and diabolical points in infinitely dimensional open
quantum systems.
\end{abstract}

\section{Introduction}
Non-Hermitian Hamiltonians, for systems with properly
balanced dissipation and amplification, have real energy spectra
if they exhibit the parity-time ($ \cal PT $) symmetry, as shown by Bender
and Boettcher~\cite{Bender1998}. That discovery has lead to the
development of non-Hermitian quantum mechanics~\cite{Bender2003,Bender2007,
El-Ganainy2018,Ashida2020} and has triggered
impressive research interest ranging from studying fundamental
aspects of quantum physics~\cite{Mostafazadeh2003,
Mostafazadeh2004,Mostafazadeh2010,Znojil2008,Brody2014,Bagarello2016} to proposing applications in quantum metrology,
optics, optomechanics, and photonics~\cite{Feng2017,El-Ganainy2019,Parto2021}.
Studies of non-Hermitian quantum mechanics include
also finding quantum analogues of general relativity concepts
(like Einstein's quantum elevator~\cite{Ju2022}), proving various
no-go theorems for information processing with non-Hermitian
systems~\cite{Znojil2016,Ju2019}, and even proposals to apply
specific non-Hermitian systems, with energy spectra corresponding to the
zeros of the Riemann zeta function, aimed at proving the Riemann
hypothesis~\cite{Bender2017}.

Recently, a considerable interest in studying non-Hermitian systems has been
focused on their exceptional points (EPs), which occur, e.g., at
the phase transitions between the $ \cal PT $ and non-$ \cal PT $~regimes. Various
applications of EPs for refined quantum control of dissipative
and/or amplified systems have been proposed (for reviews
see~\cite{Ozdemir2019,Miri2019,Parto2021} and references
therein).

Most of these studies on EPs are limited to Hamiltonian EPs, which
correspond to the degeneracies of the eigenvalues of non-Hermitian
Hamiltonians associated with their coalescent eigenvectors. We
note that these EPs may be called semiclassical, because they are
not affected by quantum jumps, as explicitly discussed
in~\cite{Minganti2019} based on the quantum-trajectory
theory~\cite{Carmichael1993}, which is also referred to as the
Monte Carlo wave-function method~\cite{Dalibard1992,Molmer1993} or
the quantum-jump approach~\cite{Plenio1998}. Indeed, non-Hermitian
Hamiltonians describe only a continuous nonunitary dissipation of
an open system, but its full quantum description requires also the
inclusion of quantum jumps in the system evolution. Recently,
these semiclassical Hamiltonian EPs have been generalized to the
quantum regime by analyzing the EPs of quantum Liouvillians
instead of those of non-Hermitian
Hamiltonians~\cite{Minganti2019}. Specifically, quantum EPs (QEPs)
can be defined as the degeneracies of the eigenvalues
corresponding to coalescent eigenmatrices (eigenoperators) of the
quantum Liouvillian superoperator for a Lindblad master equation.
Clearly, such an approach to EPs includes the effect of quantum
jumps, as it is based on the standard master-equation
approach~\cite{Breuer2007} with a trivial metric for arbitrary
systems. In contrast to this, a complete description of the
evolution of a non-Hermitian system within the Bender-Boettcher
quantum mechanics requires also calculating the evolution of a
nontrivial metric of the system. This is rather complicated and
nonintuitive, but necessary to obtain physical results without
violating any no-go theorems~\cite{Znojil2016,Ju2019}. The effects
of quantum jumps on QEPs have been recently confirmed
experimentally in~\cite{Gunderson2021,Chen2022}.

QEPs and Hamiltonian EPs are, in general, different, although they can be
equivalent for classical-like systems (e.g., linearly coupled
harmonic oscillators~\cite{Minganti2019}) or specific
finite-dimensional systems~\cite{Naghiloo2019}. Anyway, one can
experimentally observe the transition of QEPs into Hamiltonian EPs by a
proper postselection of quantum trajectories within the
hybrid-Liouvillian formalism of Ref.~\cite{Minganti2020}. We also
note that if the eigenvectors corresponding to degenerate
eigenvalues of Hamiltonians or Liouvillians do \emph{not}
coalesce, then such points are refereed to as diabolical points.
These points have also wide applications in witnessing
quantum effects including phase transitions. For example, quantum
Liouvillian diabolical points can reveal dissipative phase transitions and a
Liouvillian spectral collapse~\cite{Minganti2021a,Minganti2021b}.

For the above reasons, the determination of QEPs and their properties represents
an important task, which can lead to applications in
quantum technologies, including quantum metrology.
Unfortunately, the standard approach of finding QEPs via the
eigenvalue problem of Liouvillians becomes quite inefficient for
multi-qubit or multi-level quantum systems. For systems with infinitely-dimensional Hilbert
spaces, the determination of EPs and QEPs is even more challenging.
Here, we develop an efficient method based on the Heisenberg-Langevin equations for
finding QEPs, and we show the equivalence of QEPs found by these
two approaches. We apply the developed method for the determination of
QEPs for a system of $ M $ bosonic modes mutually
interacting via quadratic nonlinear Hamiltonians. Such
Hamiltonians are appealing as they lead to the linear
exactly-soluble Heisenberg operator equations and, simultaneously,
allow to describe nonclassical optical fields including squeezed
\cite{Luks1988} and sub-Poissonian fields \cite{Mandel1995}. Such
fields are analyzed in the framework of a generalized
superposition of a signal and noise \cite{Perina1991} that describes
quantum Gaussian fields.

The applied method uses the analysis of the dynamical equations
for the moments of field operators. It is based upon the
equivalence of the eigenfrequency analysis of the Liouvillian and
the analysis of an arbitrary complete set of operators of
measurable quantities. In the case of quadratic Hamiltonians the field operators moments
(FOMs) can be chosen as a suitable set of such
measurable operators. These moments form specific structures \cite{Arkhipov2021,Arkhipov2021a,Arkhipov2022} that
exhibit QEPs of varying (exceptional) degeneracies [QEP degeneracies] and
multiplicities [diabolical degeneracies (QDP degeneracies) for quantum hybrid diabolical exceptional points (QHPs)].
Contrary to other more general Hamiltonians,
the first-order FOMs of quadratic Hamiltonians form a closed
set of dynamical equations. Moreover, the eigenfrequencies
obtained from these equations allow for the determination of those
for higher-order FOMs. These FOMs then
provide a complete structure of eigenfrequencies whose numbers
increase with the increasing order of the moments. Among others,
this results in the identification of eigenfrequencies
with infinite QEP and QDP degeneracies forming QEPs and QHPs.
There also occur cases in which degenerate eigenfrequencies remain the same at a QEP, i.e.
this QEP is not identified by the spectrum. We may refer to hidden quantum exceptional points (\emph{hidden QEPs}).

The paper is organized as follows. The equivalence of the system
eigenfrequencies analyses based on the Liouvillian and Heisenberg equations for the operators of measurable quantities is
discussed in general in Sec.~II. Both approaches are compared in
Sec.~III by considering a simple system of a damped two-level
atom. Section~IV is devoted to the application of the Heisenberg-Langevin
equations for an $ M $-mode bosonic system with a quadratic
Hamiltonian and the eigenfrequency analysis of the equations for
FOMs. The eigenfrequencies and their general structure
are discussed in Sec.~V by considering a two-mode bosonic system
with a general quadratic Hamiltonian. Conclusions are drawn in
Sec.~VI. The correspondence between the generalized master
equation and the Heisenberg-Langevin equations is discussed in
Appendix~A.

\section{Equivalence of eigenfrequency analyses in the spaces
of statistical operators and measurable operators}

Before we address the analysis of QEPs for specific models, we show
that the eigenfrequency analysis of a Liouvillian $ \hat{\cal L} $
in the space of statistical operators can be equivalently replaced
by the eigenfrequency analysis for an arbitrary complete set of
measurable operators. The equivalence of projection operator
techniques, used for the derivation of generalized master equations
with their Liouvillians, and the set of the Heisenberg equations is
discussed in detail in \cite{Mori1965,Tokuyama1976,PerinaJr1995}.
The Liouvillian superoperator $ \hat{\cal L} \cdot \equiv -i/\hbar
[\hat{H},\cdot] $, where $ \cdot $ stands for an arbitrary
operator, $ [,] $ denotes the commutator, and $ \hbar $ means the
reduced Planck constant, is defined in terms of the overall
Hamiltonian $ \hat{H} $ involving the system and its reservoir in its complete description. It
is used to evolve the statistical operator $ \hat{\rho}(t) $
according to the Liouville equation
\begin{equation}   
 \frac{d\hat{\rho}(t)}{dt} =  \hat{\cal L} \hat{\rho}(t) .
\label{1}
\end{equation}

We consider an $ N $-dimensional Hilbert space with an arbitrary
basis $ |j\rangle $ for $ j=1,\ldots,N $. The corresponding basis
in the Liouville space of a statistical operator $ \hat{\rho} $ is
formed by vectors $ \hat{\rho}_{jk} \equiv |j\rangle\langle k| $.
They allow to express $ \hat{\rho} $ in terms of coefficients $
\rho_{jk} $ as follows:
\begin{equation}     
 \hat{\rho}(t) = \sum_{jk} \rho_{jk}(t) \hat{\rho}_{jk}.
\label{2}
\end{equation}
The Liouville equation in~(\ref{1}) transforms into the following
set of equations for the coefficients $ \rho_{jk} $:
\begin{equation}   
  \frac{ d\rho_{mn}(t)}{dt} =  \sum_{jk} {\cal L}_{jk}^{mn} \rho_{jk}(t),
\label{3}
\end{equation}
and $ {\cal L}_{jk}^{mn} = \langle m| ( \hat{\cal L}
|j\rangle\langle k| ) |n\rangle $. The diagonalization of the
matrix $ {\cal L}_{jk}^{mn} $ provides us complex eigenvalues $ -i
\Omega^\alpha $ that define eigenfrequencies $ \Omega^\alpha $.
The corresponding eigenvectors $ l_{jk}^\alpha $ form the
eigenoperators $ \hat{l}^\alpha = \sum_{jk} l_{jk}^\alpha
|j\rangle\langle k| $. It holds for an anti-Hermitian
superoperator $ \hat{\cal L} $ that for any eigenvalue $
-i\Omega_\alpha $ with eigenoperator $ \hat{l}^\alpha $ there also
exists the eigenvalue $ i\Omega_\alpha^* $ with the corresponding eigenoperator $
\hat{l}^{\dagger\alpha} $. Moreover the Liouvillians $ \hat{\cal
L} $ have one eigenfrequency $ \Omega_0 = 0 $ with the Hermitian
eigenoperator $ \hat{l}^0 = \hat{l}^{\dagger 0} $ and $ {\rm Tr}
\{ \hat{l}^0 \} = 1 $ that describes the steady state. Contrary to
$ \hat{l}^0 $, the remaining eigenoperators $ \hat{l}^\alpha $ are
non-Hermitian and obey $ {\rm Tr} \{ \hat{l}^\alpha \} = 0 $.

In parallel to Eq.~(\ref{2}), an arbitrary statistical operator $
\hat{\rho} $ can be decomposed in the basis $ \hat{l}^\alpha $ as
follows
\begin{equation}     
 \hat{\rho}(t) = \sum_{\alpha} \big[ \rho^\alpha (t) \hat{l}^\alpha
  + \rho^{\alpha *}(t) \hat{l}^{\alpha \dagger} \big] ,
\label{4}
\end{equation}
where $ \rho^\alpha(t) = {\rm Tr} \{ \hat{\rho}(t)
\hat{l}^{\alpha\dagger} \} $. The evolution of the statistical
operator $ \hat{\rho}(t) $ is then expressed along the formula
\begin{equation}     
 \hat{\rho}(t) = \sum_{\alpha} \big[ \exp(-i\Omega_\alpha t) \rho^\alpha(0) \hat{l}^\alpha
  + \exp(i\Omega_\alpha^* t) \rho^{\alpha *}(0)
  \hat{l}^{\alpha\dagger} \big]
\label{5}
\end{equation}
that `separates' the evolution for different eigenfrequencies $
\Omega_\alpha $ and $ -\Omega_\alpha^* $. The coefficients $
\rho^\alpha(0) $ in Eq.~(\ref{5}) characterize the initial state.
The mean value $ \langle A\rangle(t) $ of an arbitrary Hermitian
operator $ \hat{A}(0) $ of a measurable quantity is then
determined in the Schr\"{o}dinger picture as
\begin{eqnarray}   
 \langle A\rangle(t) \equiv {\rm Tr} \{ \hat{\rho}(t)\hat{A}(0) \}
  = \sum_{\alpha} \big[ \exp(-i\Omega_\alpha t) \rho^\alpha(0)
  A^{\alpha *}(0) 
  + \exp(i\Omega_\alpha^* t) \rho^{\alpha *}(0)
  A^{\alpha}(0) \big],
\label{6}
\end{eqnarray}
using Eq.~(\ref{5}) and the coefficients $ A^\alpha(0) = {\rm Tr}
\{ \hat{A}(0) \hat{l}^{\alpha\dagger} \} $.

Now let us consider an arbitrary Hermitian operator $ \hat{A}(t) $
of a measurable quantity and its evolution in the Heisenberg
picture according to the Heisenberg equation
\begin{equation}     
 \frac{d\hat{A}(t)}{dt} =  -\hat{\cal L} \hat{A}(t).
\label{7}
\end{equation}
The eigenoperators of the superoperator $ -{\cal L} $ coincide
with those of $ {\cal L} $ and the corresponding eigenvalues
differ by sign. Decomposing the operator $ \hat{A}(0) $ at $ t=0 $
into the basis $ \hat{l}^\alpha $,
\begin{equation}     
 \hat{A}(0) = \sum_{\alpha} \big[ A^\alpha (0) \hat{l}^\alpha
  + A^{\alpha *}(0) \hat{l}^{\alpha \dagger} \big] ,
\label{8}
\end{equation}
we may express the solution of the Heisenberg equation (\ref{7})
as follows:
\begin{equation}     
 \hat{A}(t) = \sum_{\alpha} \big[ \exp(i\Omega_\alpha t) A^\alpha(0)
 \hat{l}^{\alpha} + \exp(-i\Omega_\alpha^* t) A^{\alpha *}(0)
 \hat{l}^{\dagger\alpha} \big] .
\label{9}
\end{equation}
Using Eq.~(\ref{9}), the mean value $ \langle A\rangle(t) $ of the
operator $ \hat{A} $ in the Heisenberg picture is expressed in
terms of the coefficient $ \rho^\alpha(0) $ of the initial
statistical operator $ \hat{\rho}(0) $:
\begin{eqnarray}   
 \langle A\rangle(t) \equiv {\rm Tr} \{ \hat{\rho}(0)\hat{A}(t) \}
  = \sum_{\alpha} \big[ \exp(i\Omega_\alpha t) \rho^{\alpha *}(0)
  A^\alpha(0) 
  + \exp(-i\Omega_\alpha^* t) \rho^\alpha(0)
  A^{\alpha *}(0) \big].
\label{10}
\end{eqnarray}
Provided that the Liouvillian $ \hat{\cal L} $ is constructed
using a Hermitian Hamiltonian $ \hat{H} $ the eigenfrequencies $
\Omega^\alpha $ are real and the formulas for the mean values $
\langle A\rangle(t) $ in Eqs.~(\ref{6}) and (\ref{10}) coincide.
This means that the time evolution of the operator $ \hat{A}(t) $
is described only by the eigenfrequencies $ \Omega_\alpha $ and $
-\Omega_\alpha $ known from the evolution of the statistical
operator $ \hat{\rho} $ fulfilling Eq.~(\ref{1}). Once we
construct an arbitrary basis from the Hermitian operators $
\hat{A} $ of measurable quantities and analyze the evolution of
their mean values $ \langle A\rangle(t) $, we reveal all the
eigenfrequencies $ \Omega_\alpha $. The used basis can even be
chosen more generally involving also non-Hermitian operators $
\hat{A} $.

When deriving a master equation for the reduced statistical
operator $ \hat{\rho}^{\rm s} $, by tracing out the part of the
whole statistical operator $ \hat{\rho} $ belonging to the
reservoir, we apply the perturbation solution of the general
Liouville equation (\ref{1}) valid up to the second power of the
interaction Hamiltonian between the system and its reservoir. This
results in a new Liouvillian $ \hat{\cal L}^{\rm s} $ that has a
more general form compared to that expressed as a commutator with
the Hamiltonian $ \hat{H} $. We may diagonalize this more general
form of the system Liouvillian $ \hat{\cal L}^{\rm s} $ to reveal,
in general, complex eigenvalues $ -\Omega_\alpha $ and $
\Omega_\alpha^* $ and the accompanying eigenoperators $
\hat{l}^\alpha $ and $ \hat{l}^{\alpha \dagger} $. Alternatively,
we convert the system Liouville equation in Eq.~(\ref{1}) into a
coupled set of differential equations for the mean values $
\langle A\rangle(t) $ of operators $ \hat{A} $ that form a basis
in the space of system operators of measurable quantities.
According to Eq.~(\ref{6}), the complex eigenfrequencies of the
dynamics matrix of this set of differential equations coincide
with the eigenfrequencies revealed by a direct diagonalization of
the system Liouvillian $ \hat{\cal L}^{\rm s} $. For details, see
Appendix~A. The differential equations for the mean values $
\langle A\rangle(t) $ can alternatively be derived from a closed
set of the Heisenberg equations written for both the system
operators $ \hat{A} $ forming the basis and operators of the
reservoir. The reservoir operators can suitably be eliminated
keeping the validity of the solution up to the second power of the
interaction Hamiltonian. This leads to the Heisenberg-Langevin
equations. This method known as the Wigner-Weisskopf model of
damping \cite{Perina1991} for a bosonic mode interacting with
bosonic reservoir provides equivalent derivation of the
differential equations for the mean values $ \langle A\rangle(t)
$. This stems from the above discussed equivalence of both
descriptions in the Schr\"{o}dinger and Heisenberg pictures,
and equivalent approximations when eliminating the reservoir
degrees of freedom \cite{Vogel2006}.

The use of the Heisenberg equations for suitable operators instead
of the master equation may provide qualitative advantage compared
to the eigenfrequency analysis of the Liouvillian. We demonstrate
this by analysing two important examples: a damped two-level atom
in Sec.~III and the system of mutually interacting bosonic modes
with the quadratic interaction Hamiltonian in Secs.~IV and V.
Whereas the analysis of a damped two-level atom in a
finite-dimensional Liouville space leads to a deeper insight into
the method based on the Heisenberg equations by its comparison
with a direct diagonalization of the Liouvillian, we reveal the
structure and values of eigenfrequencies in the model of bosonic
modes relying just on the Heisenberg equations.

\section{Eigenfrequency analyses of a damped two-level atom}

We demonstrate the above discussed equivalence of both approaches
in the determination of the system eigenfrequencies by analyzing,
probably, the simplest possible physical system --- a damped
two-level atom. Its Liouville space of statistical operators has
dimension 4 and so its Liouvillian $ {\cal L} $ has 4
eigenfrequencies and eigenoperators. Moreover, as the system is
finite dimensional, powers of the operators of measurable
quantities do not play an important role in the eigenfrequency
analysis because they are just a linear superposition of operators
from any chosen basis in the space of measurable operators. The
eigenfrequency analysis performed by both approaches also results
in their detailed comparison.

The considered two-level atom has the
ground state $ |0\rangle $ and the excited state $ |1\rangle $. It is
described by the Pauli operators $ \hat{\sigma}_x = \hat{\sigma}_+
+ \hat{\sigma}_- $, $ \hat{\sigma}_y = (\hat{\sigma}_+ -
\hat{\sigma}_-)/i $ and $ \hat{\sigma}_z = \hat{\sigma}_1 -
\hat{\sigma}_0 $; $ \hat{\sigma}_0 \equiv |0\rangle\langle 0| $, $
\hat{\sigma}_1 \equiv |1\rangle\langle 1| $, $ \hat{\sigma}_+
\equiv |1\rangle\langle 0| $, and $ \hat{\sigma}_- \equiv
|0\rangle\langle 1| $. Hamiltonian $ \hat{H}^{\rm s} $ of the
two-level atom with the excitation energy $ \hbar\omega $ is
written as
\begin{equation}   
 \hat{H}^{\rm s} = \hbar\omega\hat{\sigma}_z/2.
\label{11}
\end{equation}

\subsection{Analysis via the Liouvillian superoperator}

We begin with the eigenfrequency analysis of the corresponding
Liouvillian. We assume that the atom is damped via the interaction with a
reservoir based on the $ \hat{\sigma}_x $ operator. The
corresponding Liouvillian $ \hat{\cal L}_x $ is derived in the
form
\begin{equation}   
 \hat{\cal L}_x \cdot = \gamma_x \left[ \hat{\sigma}_x \cdot
  \hat{\sigma}_x -  1 \cdot  \right],
\label{12}
\end{equation}
where the dot $ \cdot $ stands for an arbitrary operator. For a
more general form of damping and the corresponding analysis of the
Liouvillian, see \cite{Minganti2019}.

Expressing the statistical operator $ \hat{\rho}^{\rm s} $ of the
two-level atom in a suitable basis,
\begin{equation}  
 \hat{\rho}^{\rm s} = \rho_{00}^{\rm s} |0\rangle\langle 0| +
  \rho_{01}^{\rm s} |0\rangle\langle 1| + \rho_{10}^{\rm s} |1\rangle\langle 0|
  + \rho_{11}^{\rm s} |1\rangle\langle 1|,
\label{13}
\end{equation}
we transform the Liouville equation (\ref{1}) into the following
set of linear differential equations for the coefficients of the
decomposition:
\begin{eqnarray}   
 &&\frac{d}{dt} \left[ \begin{array}{c} \rho_{00}^{\rm s} \\
  \rho_{01}^{\rm s} \\ \rho_{10}^{\rm s} \\ \rho_{11}^{\rm s}
  \end{array} \right] = -i {\bf M}
    \left[ \begin{array}{c} \rho_{00}^{\rm s} \\
  \rho_{01}^{\rm s} \\ \rho_{10}^{\rm s} \\ \rho_{11}^{\rm s}
  \end{array} \right] , \label{14} \\
 && {\bf M} = \left[ \begin{array}{cccc}
    -i\gamma_x & 0 & 0 & i\gamma_x \\
    0 & -\omega - i\gamma_x & i\gamma_x & 0 \\
    0 & i\gamma_x & \omega - i\gamma_x & 0 \\
    i\gamma_x & 0 & 0 & -i\gamma_x \end{array} \right] .
\label{15}
\end{eqnarray}

The dynamics matrix $ {\bf M} $ defined in Eq.~(\ref{15}) has 4 eigenfrequencies:
\begin{eqnarray}   
 \Omega_0^{\rm s} = 0, \hspace{5mm} 
 \Omega_{1,2}^{\rm s} = \pm \Omega^{\rm s} -i\gamma_x , \hspace{5mm} 
 \Omega_{3}^{\rm s} = -2i\gamma_x,
\label{16}
\end{eqnarray}
where $ \Omega^{\rm s} = \sqrt{ \omega^2 - \gamma_x^2 } $. The
eigenfrequencies $ \Omega_{1,2}^{\rm s} $ identify a possible QEP
for
\begin{equation}   
  \omega = \gamma_x .
\label{17}
\end{equation}
We may also determine the corresponding eigenvectors that are
conveniently written as columns of the transformation matrix $
{\bf P} $:
\begin{equation}   
 {\bf P} = \frac{1}{2}\left[ \begin{array}{cccc}
    1 & 0 & 0 & 1 \\
    0 & i\gamma_x & i\gamma_x & 0 \\
    0 & \omega + \Omega^{\rm s} & \omega - \Omega^{\rm s} & 0 \\
    1 & 0 & 0 & -1 \end{array} \right] .
\label{18}
\end{equation}
We note that the first eigenvector $ \hat{\rho}_0^{\rm s} $ for $
\Omega_0^{\rm s} $ that describes the stationary state is
normalized such that $ {\rm Tr}\{\hat{\rho}_0^{\rm s}\} = 1 $. All
other eigenvectors are traceless ($ {\rm Tr}\{\hat{\rho}_j^{\rm
s}\} = 0 $ for $ j=1,2,3 $) and no specific norm is introduced to
normalize them. We have $ \Omega^{\rm s} = 0 $ for the condition
in Eq.~(\ref{17}) and so the second and third eigenvectors
(columns) in the matrix $ {\bf P} $ in Eq.~(\ref{18}) coalesce
confirming the presence of an QEP.

\subsection{Analysis via the Heisenberg-Langevin equations}

Now let us apply the second approach. To derive the appropriate
Heisenberg-Langevin equations we need to specify the system
interaction with the reservoir represented by a large group of
two-level atoms. Their Hamiltonian $ \hat{H}^{\rm r}_0 $ and a
suitable interaction Hamiltonian $ \hat{H}^{\rm r}_{\rm int} $ are
expressed as:
\begin{eqnarray}   
 \hat{H}^{\rm r}_0 = \hbar \sum_{j} \omega_j^{\rm r} \hat{\sigma}_{j,z}^{\rm r}/2, \hspace{5mm} 
 \hat{H}^{\rm r}_{\rm int} = \hbar \sum_{j} \kappa_j^{\rm r} \hat{\sigma}_{j,x}^{\rm r}
  \hat{\sigma}_x ,
\label{19}
\end{eqnarray}
where $ \omega_j^{\rm r} $ stands for the frequency
of the $ j $th reservoir two-level atom that is coupled with the
analyzed two-level atom via the real coupling constant $
\kappa_j^{\rm r} $. The Pauli operators of the reservoir two-level
atoms are introduced in analogy to those of the analyzed
system.

We may write the Heisenberg equations for both analyzed two-level
atom and reservoir two-level atoms described by the overall
Hamiltonian $ \hat{H}^{\rm s} + \hat{H}^{\rm r}_{\rm int} +
\hat{H}^{\rm r}_0 $. A systematic elimination of the reservoir
operators then results in the following Heisenberg-Langevin
equation for an arbitrary operator $ \hat{X} $ \cite{Vogel2006}:
\begin{eqnarray}   
 \frac{d\hat{X}(t)}{dt} = \frac{i}{\hbar} [ \hat{H}^{\rm s}(t),\hat{X}(t) ]
   - \gamma_x/2 [ [ \hat{X}(t),\sigma_x(t)] ,\sigma_x(t)] 
   - i [ \hat{X}(0),\sigma_x(0)]
  \hat{F}_x(t).
\label{20}
\end{eqnarray}
In Eq.~(\ref{20}) the Langevin operator force $ \hat{F}_x(t) $,
defined as
\begin{equation}    
 \hat{F}_x(t) = \exp(-i\hat{H}^{\rm r}_0 t/\hbar) \sum_{j} \kappa_j^{\rm r}
  \hat{\sigma}_{j,x}^{\rm r}(0) \exp(i\hat{H}^{\rm r}_0 t/\hbar),
\label{21}
\end{equation}
represents the back-action of the reservoir two-level atoms to the
analyzed atom. The damping constant $ \gamma_x $ is derived from
the properties of the reservoir Langevin operator force $
\hat{F}_x(t) $ along the formula:
\begin{equation} 
 \gamma_x = 2 \int_{0}^{\infty} d\tau\, \langle \hat{F}_x(\tau)\hat{F}_x(0)\rangle_{\rm r}
  \exp(-i\omega\tau) .
\label{22}
\end{equation}
Moreover, the reservoir properties imply that $ \langle
\hat{F}_x(t)\rangle = 0 $.

Using the general formula in Eq.~(\ref{20}), we write the
Heisenberg-Langevin equations for the four operators $ \hat{\sigma}_0
$, $ \hat{\sigma}_1 $, $ \hat{\sigma}_+ $, and $ \hat{\sigma}_- $
that form a basis in the space of operators of the measurable
quantities:
\begin{eqnarray}  
 \frac{d\hat{\sigma}_0(t)}{dt} = - \frac{d\hat{\sigma}_1(t)}{dt} &=& - \gamma_x \hat{\sigma}_0(t) +
  \gamma_x \hat{\sigma}_1(t) - \hat{\sigma}_y(0) \hat{F}_x(t),
  \nonumber \\
 \frac{d\hat{\sigma}_+(t)}{dt} = \left[ \frac{d\hat{\sigma}_-(t)}{dt}\right]^\dagger &=&
   i\omega \hat{\sigma}_+(t) - \gamma_x \hat{\sigma}_+(t) 
    + \gamma_x \hat{\sigma}_-(t) - i\hat{\sigma}_z(0)\hat{F}_x(t).
\label{23}
\end{eqnarray}
Applying averaging over the reservoir operators in Eq.~(\ref{23}), we arrive
at the following equations for the mean values of the system
operators:
\begin{equation}   
 \frac{d}{dt} \left[ \begin{array}{c} \langle\hat{\sigma}_0\rangle  \\
  \langle\hat{\sigma}_+\rangle  \\ \langle\hat{\sigma}_-\rangle  \\ \langle\hat{\sigma}_1\rangle
  \end{array} \right] = -i {\bf M}
    \left[ \begin{array}{c} \langle\hat{\sigma}_0\rangle  \\
  \langle\hat{\sigma}_+\rangle  \\ \langle\hat{\sigma}_-\rangle  \\ \langle\hat{\sigma}_1\rangle
  \end{array} \right] ,
\label{24}
\end{equation}
where the dynamics matrix in Eq.~(\ref{24}) coincides with that in
Eq.~(\ref{14}) because, using the representation of the
statistical operator $ \hat{\rho} $ in Eq.~(\ref{13}), we have $
\langle\hat{\sigma}_0\rangle = \rho_{00}^{\rm s} $, $
\langle\hat{\sigma}_+\rangle = \rho_{01}^{\rm s} $, $
\langle\hat{\sigma}_-\rangle = \rho_{10}^{\rm s} $, and $
\langle\hat{\sigma}_1\rangle = \rho_{11}^{\rm s} $. Thus, the
eigenfrequencies obtained by both approaches are identical.

We may also exploit the transformation matrix $ {\bf P} $ in
Eq.~(\ref{18}) to express the evolution of the mean values of the
above operators:
\begin{eqnarray} 
 \left[ \begin{array}{c} \langle\hat{\sigma}_0\rangle(t)  \\
  \langle\hat{\sigma}_1\rangle(t)
  \end{array} \right] &=& \exp(-\gamma_xt)
  \left[ \begin{array}{cc} {\rm ch}(\gamma_xt)
  & {\rm sh}(\gamma_xt) \\ {\rm sh}(\gamma_xt) & {\rm ch}(\gamma_xt)
  \end{array} \right] \left[ \begin{array}{c} \langle\hat{\sigma}_0\rangle(0)  \\
  \langle\hat{\sigma}_1\rangle(0) \end{array} \right] , \nonumber \\
  \left[ \begin{array}{c} \langle\hat{\sigma}_+\rangle(t)  \\
  \langle\hat{\sigma}_-\rangle(t)
  \end{array} \right] &=& \exp(-\gamma_xt)
   \left[ \begin{array}{cc} {\rm c}(t) + i\frac{\omega}{\Omega^{\rm s}}
   {\rm s}(t) & \frac{\gamma_x}{\Omega^{\rm s}} {\rm s}(t) \\
   \frac{\gamma_x}{\Omega^{\rm s}} {\rm s}(t) & {\rm c}(t) - i\frac{\omega}{\Omega^{\rm s}}
   {\rm s}(t) \end{array} \right] 
   \left[ \begin{array}{c} \langle\hat{\sigma}_+\rangle(0)  \\
  \langle\hat{\sigma}_-\rangle(0) \end{array} \right],
\label{25}
\end{eqnarray}
where $ {\rm c}(t) = \cos(\Omega^{\rm s}t) $, $ {\rm s}(t) =
\sin(\Omega^{\rm s}t) $, and $ \Omega^{\rm s} = \sqrt{ \omega^2 -
\gamma_x^2 } $; $ {\rm sh} $ and $ {\rm ch} $ denote the
hyperbolic sine and cosine functions. We can see from the solution
in Eq.~(\ref{25}) that there exist two subspaces in the space of
the operators of measurable quantities whose dynamics differ.
Whereas there occurs only damped dynamics in the subspace spanned
by operators $ \hat{\sigma}_0 $ and $ \hat{\sigma}_1 $, the
oscillatory evolution at frequency $ \Omega^{\rm s} $ in the
subspace spanned by vectors $ \hat{\sigma}_+ $ and $
\hat{\sigma}_- $ allows to identify an QEP ($ \Omega^{\rm s} = 0
$). We may conclude in general that to identify a QEP we need to
follow the dynamics of an operator that has a nonzero overlap with
the subspace spanned by $ \hat{\sigma}_+ $ and $ \hat{\sigma}_- $.

For a real atom, monitoring the dynamics of the mean values $
\langle\hat{\sigma}_+\rangle(t) $ and $
\langle\hat{\sigma}_-\rangle(t) $ via the observation of field
polarization generated by the atom is a natural choice. It can be
accomplished by measuring, e.g., optical nutation or free
induction decay \cite{Meystre2007}. The presence of a QEP can also
be confirmed by measuring the amplitude spectra of the field
emitted from the atom or, in case of continuous-wave fields, by
spectral analysis of field intensity correlation functions
\cite{Perina1985,Arkhipov2020}.

We note that the consideration of powers of operators in any basis
in finite-dimensional spaces does not bring advantage into the
eigenfrequency analysis, because these operators can be expressed
as linear combinations of the operators from this basis. This,
among others, also implies that only the first- and second-order
correlation functions of the reservoir operator forces $ \hat{F} $
are needed. This contrasts with the behavior of systems defined in
the infinitely dimensional Liouville spaces in which moments of
operators are the most useful, as we can see below in Sec.~IV.

\section{Eigenfrequency analysis of an $ M $-mode bosonic system}

We apply the above discussed equivalence of eigenfrequency
analyses in the Schr\"{o}dinger and the Heisenberg pictures to
discuss the system of the eigenfrequencies of $ M $ mutually
interacting bosonic modes described by their annihilation ($
\hat{a}_j $, $ j=1,\ldots M $) and creation ($ \hat{a}_j^\dagger
$) operators. To avoid approximations, when solving nonlinear
differential equations, we assume that the $ M $-mode system is
described by the general quadratic Hamiltonian $ \hat{H}_0 $:
\begin{eqnarray}   
 \hat{H}_0 &=& \hbar \sum_{j,k=1}^M \mbox{\boldmath $ \epsilon $}_{jk} \hat{a}^\dagger_j
   \hat{a}_k + \hbar \sum_{j,k=1}^M \left( \mbox{\boldmath $ \kappa $}_{jk} \hat{a}_j \hat{a}_k
   + {\rm H.c.} \right),
\label{26}
\end{eqnarray}
where the elements of matrix {\boldmath $ \epsilon $}
({\boldmath $ \epsilon $}$ {}_{ij}^* $ = {\boldmath $ \epsilon $}$
{}_{ji} $) describe the linear coupling between pairs of modes,
whereas the elements of the matrix {\boldmath $ \kappa $}
characterize the nonlinear interactions between pairs of modes
(the annihilation and creation of photon pairs). Such Hamiltonian
allows to describe both squeezed-light generation and production
of entangled states. Symbol $ {\rm H.c.} $ in Eq.~(\ref{26})
replaces the Hermitian conjugated term.

The modes may be either damped or amplified. The system
Liouvillian $ \hat{\cal L}_0  \equiv -i/\hbar [\hat{H}_0, \cdot] $
has to be extended by the terms $ \hat{\cal L}_j^{\rm d} $,
\begin{equation}   
 \hat{\cal L}_j^{\rm d} \cdot = \frac{\gamma_j^{\rm d}}{2} \left[ 2\hat{a}_j \cdot
  \hat{a}_j^\dagger - \hat{a}_j^\dagger \hat{a}_j \cdot - \cdot \hat{a}_j^\dagger \hat{a}_j
  \right],
\label{27}
\end{equation}
provided that the $ j $th mode is damped with damping constant $
\gamma_j^{\rm d} $. On the other hand, the amplification of the
mode $ k $ is described by the following additional terms $
\hat{\cal L}_k^{\rm a} $,
\begin{equation}   
 \hat{\cal L}_k^{\rm a} \cdot = \frac{\gamma_k^{\rm a}}{2} \left[ 2\hat{a}_k^\dagger \cdot
  \hat{a}_k - \hat{a}_k \hat{a}_k^\dagger \cdot - \cdot \hat{a}_k \hat{a}_k^\dagger
  \right],
\label{28}
\end{equation}
using the amplification constant $ \gamma_k^{\rm a} $.

The master equation for the statistical operator $ \hat{\rho}(t) $
of the $ M $-mode bosonic system comprising the Liouvillians $ \hat{\cal L}_0 $,
$ \hat{\cal L}_j^{\rm d} $ and $ \hat{\cal L}_k^{\rm a} $
is equivalently replaced \cite{Vogel2006,Perina1991} by the
following system of the Heisenberg-Langevin equations conveniently
written in the matrix form:
\begin{equation}   
 \frac{d\hat{\bf a}(t)}{dt} = -i{\bf M}_\Omega \hat{\bf a}(t) + \hat{\bf
   L}(t) ,
\label{29}
\end{equation}
where
\begin{equation} 
  \hat{\bf a}(t) \equiv \left[ \begin{array}{c} \hat{a}_1(t) \\
   \hat{a}_1^\dagger(t) \\ \vdots \\ \hat{a}_M(t) \\
   \hat{a}_M^\dagger(t) \end{array} \right] , \hspace{5mm}
  \hat{\bf L}(t) \equiv \left[ \begin{array}{c} \hat{L}_1(t) \\
   \hat{L}_1^\dagger(t) \\ \vdots \\ \hat{L}_M(t) \\
   \hat{L}_M^\dagger(t) \end{array} \right] .
\label{30}
\end{equation}
The dynamics matrix $ {\bf M}_\Omega $ is derived from the
Hamiltonian $ \hat{H}_0 $ in Eq.~(\ref{26}) using the canonical
commutation relations. Moreover, for the $ j $th damped mode it
includes additional terms $ -\gamma_j^{\rm d} \hat{a}_j(t)/2 $ and
$ -\gamma_j^{\rm d} \hat{a}_j^\dagger(t)/2 $ on the r.h.s. of
equations for $ d\hat{a}_j(t)/dt $ and $ d\hat{a}_j^\dagger(t)/dt
$, respectively. The stochastic Langevin operator forces $
\hat{L}_j $ and $ \hat{L}_j^\dagger $ obey the relations $ \langle
\hat{L}_j(t) \hat{L}_j^\dagger(t') \rangle = \gamma_j^{\rm
d}\delta(t-t') $ and $ \langle \hat{L}_j^\dagger(t) \hat{L}_j(t')
\rangle =  0 $ in this case.
Similarly, concerning the $ k $th amplified mode the dynamics
matrix $ {\bf M}_\Omega $ contains the terms $ \gamma_k^{\rm a}
\hat{a}_k(t) /2 $  and $ \gamma_k^{\rm a}\hat{a}_k^\dagger(t)/2 $
on the r.h.s. of equations for $ d\hat{a}_k(t)/dt $ and $
d\hat{a}_k^\dagger(t)/dt $, respectively. The relations $ \langle
\hat{L}_k(t) \hat{L}_k^\dagger(t') \rangle = 0 $ and $ \langle
\hat{L}_k^\dagger(t) \hat{L}_k(t') \rangle = \gamma_k^{\rm
a}\delta(t-t') $ are obtained in this case. The Dirac $ \delta $
functions characterizing the temporal correlations of Langevin
forces reflect the Markovian character of the interaction with
spectrally broadband reservoirs in individual modes
\cite{Meystre2007}. More details are given in Appendix~A.

The frequency analysis of the dynamics matrices of the differential
equations for the FOMs allows us to
completely determine all the eigenfrequencies of the Liouvillian $
\hat{\cal L} $ and to reveal their structure. This is owing to the
linear form of the corresponding Heisenberg-Langevin equations in
Eq.~(\ref{30}).

We first transform the original field operators $ \hat{a}_j $ and
$ \hat{a}_j^\dagger $, grouped in the vector $ {\bf a} $, to
field operators $ \hat{b}_j $ and $ \hat{b}_j^\dagger $ forming
the vector $ {\bf b}^{\rm T} = (\hat{b}_1,\hat{b}_1^\dagger,
\ldots, \hat{b}_M,\hat{b}_M^\dagger) $ via a suitable
transformation $ {\bf P} $ such that the resulting
Heisenberg-Langevin equations have a diagonal dynamics matrix $
{\bf \Omega} = {\rm diag} \{ \Omega_1, -\Omega_1^*, \ldots,
\Omega_M, -\Omega_M^* \} $:
\begin{equation}   
 \frac{d\hat{\bf b}(t)}{dt} = -i{\bf \Omega} \hat{\bf b}(t) + \hat{\bf
   K}(t) ,
\label{31}
\end{equation}
where
\begin{equation} 
 {\bf \Omega} = {\bf P}^{-1} {\bf M}_\Omega {\bf
  P}, \quad  \hat{\bf b}(t) = {\bf P}^{-1} \hat{\bf a}(t),
  \quad \hat{\bf K}(t) = {\bf P}^{-1} \hat{\bf L}(t) .
\label{32}
\end{equation}
The stochastic Langevin operator forces, which are grouped in the
vector $ \hat{\bf K}^{\rm T} = (\hat{K}_1,\hat{K}_1^\dagger,
\ldots, $ $ \hat{K}_M,\hat{K}_M^\dagger) $, induce their Gaussian
and Markovian character from the original stochastic operator
Langevin forces, as written in the vector $ \hat{\bf L} $. The set
of equations in Eq.~(\ref{31}) can be solved as follows:
\begin{eqnarray}   
 \hat{\bf b}(t) = \exp(-i{\bf \Omega}t) \hat{\bf b}(0) 
  + \int_{0}^{t} dt' \exp[-i{\bf \Omega}(t-t')] \hat{\bf K}(t') .
\label{33}
\end{eqnarray}

Now let us analyze the equations for FOMs step by step by
increasing the order of FOMs. The analysis of equations for the
first-order FOMs written for either the original operators,
\begin{equation}   
 \frac{d\langle \hat{\bf a}\rangle (t)}{dt} = -i{\bf M}_\Omega \langle \hat{\bf a}\rangle(t),
\label{34}
\end{equation}
or the transformed operators,
\begin{equation}   
 \frac{d\langle \hat{\bf b}\rangle (t)}{dt} = -i{\bf \Omega} \langle \hat{\bf b}\rangle(t),
\label{35}
\end{equation}
immediately gives us the basic set of eigenfrequencies contained
on the diagonal of the matrix $ {\bf \Omega} $.

Combining Eqs.~(\ref{31}) and (\ref{33}), we arrive at the
following differential equations for the second-order FOMs (for the case without the Langevin forces, see
\cite{Arkhipov2021}):
\begin{eqnarray}   
 \frac{d\langle \hat{\bf b}_k \hat{\bf b}_l\rangle (t)}{dt} &=& -i {\bf \Omega}_{kl}^{(2)}
   \langle \hat{\bf b}_k \hat{\bf b}_l \rangle(t) + \tilde{\bf K}_{kl} ,
\label{36} \\
 {\bf \Omega}_{kl}^{(2)} &=& {\bf \Omega}_{kk} + {\bf \Omega}_{ll} ,
\label{37}
\end{eqnarray}
in which the matrix $ \tilde{\bf K} $ contains time-independent
second-order correlation functions of the Langevin forces written
in vector $ \hat{\bf K} $:
\begin{equation}    
 \tilde{\bf K}_{kl}\delta(t-t') =  \langle \hat{\bf K}_k(t) \hat{\bf K}_l(t') \rangle .
\label{38}
\end{equation}
The solution of the `diagonal' form of the equations for the
second-order FOMs, written in Eq.~(\ref{36}), is
obtained in the form similar to that of Eq.~(\ref{33}). This
reveals that the emerging eigenfrequencies $ {\bf \Omega}_{kk} + {\bf
\Omega}_{ll} $ can be expressed in terms of the sums of those from
the basic set.

The structure of differential equations for the third-order FOMs
is more general, as it contains also the time-dependent
first-order FOMs on their r.h.s.:
\begin{eqnarray}   
 \frac{d\langle \hat{\bf b}_k \hat{\bf b}_l \hat{\bf b}_m\rangle (t)}{dt} &=& -i {\bf \Omega}_{klm}^{(3)}
  \langle \hat{\bf b}_k \hat{\bf b}_l \hat{\bf b}_m \rangle(t) 
    + \tilde{\bf K}_{kl} \langle \hat{\bf b}_m \rangle(t)  + \tilde{\bf K}_{km} \langle \hat{\bf b}_l \rangle(t)
    + \tilde{\bf K}_{lm} \langle \hat{\bf b}_k \rangle(t),
\label{39} \\
 {\bf \Omega}_{klm}^{(3)} &=& {\bf \Omega}_{kk} + {\bf \Omega}_{ll}
  +{\bf \Omega}_{mm}.
\label{40}
\end{eqnarray}
The solution to the set of equations in Eq.~(\ref{39}) can again
be expressed in the form of Eq.~(\ref{33}). Whereas the
homogeneous part of the solution to Eq.~(\ref{39}) oscillates at
frequencies $ {\bf \Omega}_{kk} + {\bf \Omega}_{ll} +{\bf
\Omega}_{mm} $, the nonhomogeneous part of the solution contains
additional frequencies $ {\bf \Omega}_{nn} $, as it can be checked
by a direct calculation.

The general structure of the differential equations for the fourth-
and higher-order FOMs is the same as that for
the third-order FOMs. To demonstrate this, we
derive the following equations for the fourth-order FOMs:
\begin{eqnarray}   
 \frac{d\langle \hat{\bf b}_k \hat{\bf b}_l \hat{\bf b}_m \hat{\bf b}_n\rangle (t)}{dt} &=&
  -i  {\bf \Omega}_{klmn}^{(4)} \langle \hat{\bf b}_k \hat{\bf b}_l \hat{\bf b}_m  \hat{\bf b}_n \rangle(t) 
  + \tilde{\bf K}_{kl} \langle \hat{\bf b}_m \hat{\bf b}_n  \rangle(t)  + \tilde{\bf K}_{km} \langle \hat{\bf
b}_l \hat{\bf b}_n \rangle(t) \nonumber \\ 
 & &  \hspace{-15mm} + \tilde{\bf K}_{kn} \langle \hat{\bf b}_l \hat{\bf b}_m\rangle(t)  + \tilde{\bf K}_{lm} \langle \hat{\bf b}_k \hat{\bf b}_n\rangle(t)
   + \tilde{\bf K}_{ln} \langle \hat{\bf b}_k \hat{\bf b}_m\rangle(t) + \tilde{\bf K}_{mn} \langle \hat{\bf b}_k \hat{\bf b}_l\rangle(t) ,
\label{41} \\
  {\bf \Omega}_{klmn}^{(4)} &=& {\bf \Omega}_{kk} + {\bf \Omega}_{ll}  +{\bf \Omega}_{mm} +{\bf \Omega}_{nn}.
\label{42}
\end{eqnarray}
The solution to Eq.~(\ref{41}) in the form of Eq.~(\ref{33})
reveals the two types of eigenfrequencies: $ {\bf \Omega}_{kk} +
{\bf \Omega}_{ll} +{\bf \Omega}_{mm} +{\bf \Omega}_{nn} $ and $
{\bf \Omega}_{kk} + {\bf \Omega}_{ll} $.

In general, the analysis of differential equations for $ p $th-order FOMs (for $ p > 3 $) reveals the sums of
$ p $ (for the homogeneous solution) and $ p-2 $ (for the nonhomogeneous
solution) eigenfrequencies from the basic set.

The analysis of differential equations for the FOMs of increasing order gradually reveals all the eigenfrequencies
and their degeneracies. At a general level, we may determine the
number $ n_\Omega^{(p)} $ of different eigenfrequencies provided by the
analysis of the homogeneous solution of the set of equations for $
p $th-order FOMs. This number is given by the number
of independent $ p $th-order moments. The overall number of $ p
$th-order moments equals $ (2M)^p $. However, the mean values of
the products of commuting field operators are insensitive to
their ordering [e.g., $ \langle \hat{a}_1\hat{a}_2\rangle = \langle
\hat{a}_2\hat{a}_1\rangle $ for the field operators $ \hat{a}_1 $ and $
\hat{a}_2 $]. Also, if two field operators do not commute, the
mean values of the products of $ p $ operators with different
ordering of noncommuting operators differ by the mean value of
the product of $ p-2  $ operators [e.g., $ \langle \hat{x}
\hat{a}\hat{a}^\dagger \hat{y} \rangle = \langle
\hat{x}\hat{a}^\dagger\hat{a} \hat{y} \rangle + \langle \hat{x}
\hat{y}\rangle $ for the field operators $ \hat{a} $ and $
\hat{a}^\dagger $ and arbitrary operators $ \hat{x} $ and $
\hat{y} $]. Such lower-order FOMs form the
r.h.s. of the differential equations for the $ p $th-order FOMs, together with the terms arising from the
interaction with the reservoirs.

The numbers of independent FOMs and, thus, the numbers $
n_\Omega^{(p)} $ of eigenfrequencies can be easily determined
using the combinations of numbers. Considering the moments up to
the fourth order, we arrive at the following formulas:
\begin{eqnarray}    
 n_\Omega^{(1)} &=& 2M , \nonumber \\
 n_\Omega^{(2)} &=& 2M + C(2M,2), \nonumber \\
 n_\Omega^{(3)} &=& 2M + 2C(2M,2) + C(2M,3), \nonumber \\
 n_\Omega^{(4)} &=& 2M + 3C(2M,2) + 3C(2M,3) + C(2M,4), 
\label{43}
\end{eqnarray}
where $ C(k,l) = k!/[(k-l)! l!] $ is the binomial coefficient.

The eigenfrequencies arising from the equations for the FOMs of different orders form specific structures.
Their analysis then allows to identify QEPs and QHPs and their
degeneracies. In the following section, we study this structure
for a system composed of $ M=2 $ modes and exhibiting both damping
and amplification.

The eigenvalue analysis of the dynamics matrices of different FOMs
provides also the corresponding eigenvectors. Whereas the obtained
eigenvalues coincide with those revealed by the eigenvalue
analysis of the corresponding Liouvillian $ {\cal L} $, the
obtained eigenvectors do not allow to construct the eigenvectors
of the Liouvillian $ {\cal L} $. The reason is that the
appropriate equations are of a different kind. Whereas the
influence of the reservoir noise is involved in the set of
equations for second-order FOMs, in Eq.~(\ref{36}), via its
nonhomogeneous solution, it is directly embedded in the form of
the Liouvillian $ {\cal L} $ whose eigenvalue problem is solved.
Also, the presence of reservoir noise leads to the coupling of
equations for the odd [and similarly even] orders of FOMs, as
shown in Eq.~(\ref{39}) [Eq.~(\ref{41})]. However and most
importantly, this coupling is specific and, as discussed above,
keeps the eigenvalues obtained for the homogenenous solutions
unchanged.

\begin{table}[ht]   
\begin{center}
\begin{tabular}{|c|c|c|c|c|}
 \hline
  $ {\bf \Omega^{\rm i}_{\rm gen}} $ & $ {\bf \Omega^{\rm r}_{\rm gen}} $ & Moments & Moment & Partial \\
   &  &  & deg.  & QDP x \\
   &  &  &  & QEP deg.\\
 \hline
 \hline
 $ \Omega^{\rm i} $ & $ \pm \Omega_1^{\rm r} $ & $ \langle \hat{b}_1 \rangle $, $ \langle \hat{b}_1^\dagger \rangle $ & 1 & 1x2\\
 \hline
  $ 2\Omega^{\rm i} $ & $ \Omega_2^{\rm r} \pm \Omega_1^{\rm r} $ & $ \langle \hat{b}_1 \hat{b}_2 \rangle $, $ \langle \hat{b}_1^\dagger \hat{b}_2 \rangle $ & 2 & 2x2 \\
 \cline{2-5}
     & $ -\Omega_2^{\rm r} \pm \Omega_1^{\rm r} $ & $ \langle \hat{b}_1 \hat{b}_2^\dagger \rangle $, $ \langle \hat{b}_1^\dagger \hat{b}_2^\dagger \rangle $ & 2 & 2x2 \\
 \cline{2-5}
   & $ \Omega_1^{\rm r} - \Omega_1^{\rm r} $ & $ \langle \hat{b}_1^\dagger \hat{b}_1 \rangle $ & 2 & 1x4  \\
 \cline{2-4}
     & $ \pm 2\Omega_1^{\rm r} $ & $ \langle \hat{b}_1^2 \rangle $, $ \langle \hat{b}_1^{\dagger 2} \rangle $ & 1 & \\
 \hline
  $ 3\Omega^{\rm i} $ & $ \pm \Omega_1^{\rm r} $  & $ \langle \hat{b}_1 \hat{b}_2 \hat{b}_2^\dagger \rangle $, $ \langle \hat{b}_1^\dagger \hat{b}_2^\dagger \hat{b}_2 \rangle $ & 6 & 6x2 \\
 \cline{2-5}
 & $ 2\Omega_2^{\rm r} \pm \Omega_1^{\rm r} $ & $ \langle \hat{b}_1 \hat{b}_2^2 \rangle $, $ \langle \hat{b}_1^\dagger \hat{b}_2^2 \rangle $ & 3 & 3x2 \\
 \cline{2-5}
 & $ -2\Omega_2^{\rm r} \pm \Omega_1^{\rm r} $ & $ \langle \hat{b}_1 \hat{b}_2^{\dagger 2} \rangle $, $ \langle \hat{b}_1^\dagger \hat{b}_2^{\dagger 2} \rangle $ & 3 & 3x2 \\
 \cline{2-5}
 & $ \Omega_2^{\rm r} + \Omega_1^{\rm r} - \Omega_1^{\rm r} $ & $ \langle \hat{b}_1^\dagger \hat{b}_1 \hat{b}_2 \rangle $ & 6 & 3x4\\
 \cline{2-4}
 & $ \Omega_2^{\rm r} \pm 2\Omega_1^{\rm r} $ & $ \langle \hat{b}_1^2 \hat{b}_2 \rangle $, $ \langle \hat{b}_1^{\dagger 2} \hat{b}_2 \rangle $ & 3 & \\
 \cline{2-5}
  & $ -\Omega_2^{\rm r} + \Omega_1^{\rm r} - \Omega_1^{\rm r} $ & $ \langle \hat{b}_1^\dagger \hat{b}_1 \hat{b}_2^\dagger \rangle $ & 6 & 3x4\\
 \cline{3-4}
  & $ -\Omega_2^{\rm r} \pm 2\Omega_1^{\rm r} $ & $ \langle \hat{b}_1^2 \hat{b}_2^\dagger \rangle $, $ \langle \hat{b}_1^{\dagger 2} \hat{b}_2^\dagger \rangle $ & 3 & \\
 \cline{2-5}
  & $ \Omega_1^{\rm r} - \Omega_1^{\rm r} \pm \Omega_1^{\rm r} $ & $ \langle \hat{b}_1^2 \hat{b}_1^\dagger \rangle $, $ \langle \hat{b}_1^{\dagger 2} \hat{b}_1 \rangle $ & 3 & 1x8\\
 \cline{2-4}
  & $ \pm 3\Omega_1^{\rm r} $ & $ \langle \hat{b}_1^3 \rangle $, $ \langle \hat{b}_1^{\dagger 3} \rangle $ & 1 &  \\
\hline
  $ 4\Omega^{\rm i} $ & $  \Omega_2^{\rm r} \pm \Omega_1^{\rm r} $  & $ \langle \hat{b}_1 \hat{b}_2^\dagger \hat{b}_2^2 \rangle $, $ \langle \hat{b}_1^\dagger \hat{b}_2^\dagger \hat{b}_2^2 \rangle $ & 12 & 12x2 \\
 \cline{2-5}
 & $ -\Omega_2^{\rm r} \pm \Omega_1^{\rm r} $  & $ \langle \hat{b}_1 \hat{b}_2^{\dagger 2} \hat{b}_2 \rangle $, $ \langle \hat{b}_1^\dagger \hat{b}_2^{\dagger 2} \hat{b}_2 \rangle $ & 12 & 12x2 \\
 \cline{2-5}
   & $  3\Omega_2^{\rm r} \pm \Omega_1^{\rm r} $ & $ \langle \hat{b}_1 \hat{b}_2^3 \rangle $, $ \langle \hat{b}_1^{\dagger} \hat{b}_2^3 \rangle $ & 4 & 4x2  \\
 \cline{2-5}
  & $  -3\Omega_2^{\rm r} \pm \Omega_1^{\rm r} $ & $ \langle \hat{b}_1 \hat{b}_2^{\dagger 3} \rangle $, $ \langle \hat{b}_1^{\dagger} \hat{b}_2^{\dagger 3} \rangle $ & 4 & 4x2 \\
 \cline{2-5}
 & $ \Omega_1^{\rm r} - \Omega_1^{\rm r} $ & $ \langle \hat{b}_1^\dagger \hat{b}_1 \hat{b}_2^\dagger \hat{b}_2 \rangle $ & 24 & 12x4 \\
 \cline{2-4}
  &  $ \pm 2\Omega_1^{\rm r} $ & $ \langle \hat{b}_1^2 \hat{b}_2^\dagger \hat{b}_2 \rangle $, $ \langle \hat{b}_1^{\dagger 2} \hat{b}_2^\dagger \hat{b}_2 \rangle $ & 12 & \\
 \cline{2-5}
   & $ 2\Omega_2^{\rm r} + \Omega_1^{\rm r} - \Omega_1^{\rm r}$ & $ \langle \hat{b}_1^\dagger \hat{b}_1 \hat{b}_2^2 \rangle $ & 12 & 6x4 \\
 \cline{2-4}
   & $ 2\Omega_2^{\rm r} \pm 2\Omega_1^{\rm r} $ & $ \langle \hat{b}_1^2 \hat{b}_2^2 \rangle $, $ \langle \hat{b}_1^{\dagger 2} \hat{b}_2^2 \rangle $ & 6 & \\
 \cline{2-5}
   & $ - 2\Omega_2^{\rm r} +\Omega_1^{\rm r} -\Omega_1^{\rm r}  $ & $ \langle \hat{b}_1^\dagger \hat{b}_1 \hat{b}_2^{\dagger 2} \rangle $ & 12 & 6x4 \\
 \cline{2-4}
   & $ -2\Omega_2^{\rm r} \pm 2\Omega_1^{\rm r} $ & $ \langle \hat{b}_1^2 \hat{b}_2^{\dagger 2} \rangle $, $ \langle \hat{b}_1^{\dagger 2} \hat{b}_2^{\dagger 2} \rangle $ & 6 &\\
 \cline{2-5}
  & $ \Omega_2^{\rm r} + \Omega_1^{\rm r} - \Omega_1^{\rm r} \pm \Omega_1^{\rm r} $ & $ \langle \hat{b}_1^\dagger \hat{b}_1^2 \hat{b}_2 \rangle $, $ \langle \hat{b}_1^{\dagger 2} \hat{b}_1 \hat{b}_2 \rangle $ & 12 &  4x8 \\
 \cline{2-4}
   & $ \Omega_2^{\rm r} \pm 3\Omega_1^{\rm r} $ & $ \langle \hat{b}_1^3 \hat{b}_2 \rangle $, $ \langle \hat{b}_1^{\dagger 3} \hat{b}_2 \rangle $ & 4 &\\
  \cline{2-5}
  & $ - \Omega_2^{\rm r} + \Omega_1^{\rm r} - \Omega_1^{\rm r} \pm \Omega_1^{\rm r} $ & $ \langle \hat{b}_1^\dagger \hat{b}_1^2 \hat{b}_2^\dagger \rangle $, $ \langle \hat{b}_1^{\dagger 2} \hat{b}_1 \hat{b}_2^\dagger \rangle $ & 12 & 4x8 \\
  \cline{2-4}
   & $ -\Omega_2^{\rm r} \pm 3\Omega_1^{\rm r} $ & $ \langle \hat{b}_1^3 \hat{b}_2^\dagger \rangle $, $ \langle \hat{b}_1^{\dagger 3} \hat{b}_2^\dagger \rangle $ & 4 & \\
 \cline{2-5}
  & $ 2\Omega_1^{\rm r} - 2\Omega_1^{\rm r} $  & $ \langle \hat{b}_1^{\dagger 2} \hat{b}_1^2 \rangle $ & 6 & 1x16 \\
 \cline{2-4}
  & $ + \Omega_1^{\rm r} - \Omega_1^{\rm r} \pm 2\Omega_1^{\rm r} $ & $ \langle \hat{b}_1^\dagger \hat{b}_1^3 \rangle $, $ \langle \hat{b}_1^{\dagger 3} \hat{b}_1 \rangle $ & 4 & \\
 \cline{2-4}
   & $ \pm 4\Omega_1^{\rm r} $ & $ \langle \hat{b}_1^4 \rangle $, $ \langle \hat{b}_1^{\dagger 4} \rangle $ & 1 & \\
 \hline
\end{tabular} 
\end{center}
 \caption{Real and imaginary parts of the complex eigenfrequencies $ \Omega_{\rm gen}^{\rm r} - i
  \Omega_{\rm gen}^{\rm i} $ derived from the equations for the FOMs up to
  fourth order which \emph{reveal} QEPs and QHPs for $ g \ne 0 $. The corresponding moments in the
  `diagonalized' field operators are written together with their degeneracy coming from
  different orderings of field operators. QDP degeneracy of QHPs (partial QDP degeneracy) derived from
  the indicated FOMs and QEP degeneracy of the constituting QEPs are given.}
\label{tab1}
\end{table}

\clearpage

\section{Spectral eigenfrequencies of a two-mode system with damping and amplification}

In this section, we analyze the system of two interacting modes:
one exhibiting damping ($ \gamma_1^{\rm d}>0 $), the other being
amplified ($ \gamma_2^{\rm a}>0 $). We consider both linear
exchange of photons between the modes (real $ \epsilon $), as well
as emission and annihilation of photon pairs in these modes.
Photon pairs can be annihilated and created either inside both
modes (real $ g $) or their photons belong to different modes
(real $ \kappa $). The corresponding Hamiltonian $ \hat{H}_0 $ is
written as follows:
\begin{eqnarray} 
 \hat{H}_0 = ( \hbar \epsilon \hat{a}^\dagger_1 \hat{a}_2 + {\rm H.c.})
  + (\hbar \kappa \hat{a}_1\hat{a}_2 + {\rm H.c.})  
  + \frac{1}{2} \sum_{j=1,2} (\hbar g\hat{a}_j^{\dagger 2}  + {\rm H.c.}) ,
\label{44}
\end{eqnarray}
which leads to the following Heisenberg-Langevin equations:
\begin{eqnarray}   
 && \frac{d}{dt} \left[ \begin{array}{c} \hat{a}_1(t) \\
  \hat{a}_1^\dagger(t) \\ \hat{a}_2(t) \\ \hat{a}_2^\dagger(t)
  \end{array} \right] = -i \tilde{\bf M}
   \left[ \begin{array}{c} \hat{a}_1(t) \\
  \hat{a}_1^\dagger(t) \\ \hat{a}_2(t) \\ \hat{a}_2^\dagger(t)
  \end{array} \right] + \left[ \begin{array}{c} \hat{L}_1(t) \\
  \hat{L}_1^\dagger(t) \\ \hat{L}_2(t) \\ \hat{L}_2^\dagger(t)
  \end{array} \right] ,
\label{45} \\
 & & \tilde{\bf M} = \left[ \begin{array}{cccc}
    -i\gamma_1^{\rm d}/2 & g & \epsilon & \kappa \\
    -g & -i\gamma_1^{\rm d}/2 & -\kappa & -\epsilon\\
    \epsilon & \kappa & i\gamma_2^{\rm a}/2 & g \\
    -\kappa & -\epsilon & -g & i\gamma_2^{\rm a}/2 \end{array} \right].
\label{46}
\end{eqnarray}
The only nonzero second-order correlation functions of the
stochastic Langevin operator forces $ \hat{L}_{1} $, $
\hat{L}_{1}^\dagger $, $ \hat{L}_{2} $, and $ \hat{L}_{2}^\dagger
$ are
\begin{eqnarray}   
 \langle \hat{L}_1(t) \hat{L}_1^\dagger(t') \rangle =
  \gamma_1^{\rm d}\delta(t-t'), \hspace{5mm} 
  \langle \hat{L}_2^\dagger(t) \hat{L}_2(t')\rangle =
  \gamma_2^{\rm a}\delta(t-t').
\label{47}
\end{eqnarray}

The diagonalization of the dynamics matrix $ \tilde{\bf M} $ in Eq.~(\ref{46}) reveals
four eigenfrequencies from the basic set:
\begin{eqnarray}   
 && {\bf \Omega} = {\rm diag}\{\Omega_1^{\rm r},-\Omega_1^{\rm r},\Omega_2^{\rm r},-\Omega_2^{\rm r}\}
     -i\Omega^{\rm i}{\rm diag}\{1,1,1,1\} ,
\label{48} \\
 && \Omega_{1,2}^{\rm r} = \sqrt{ \beta^2 - g^2 \pm
    2g\sqrt{\kappa^2+\gamma_+^2 } }, \nonumber \\
 && \Omega^{\rm i} = \gamma_-  , \nonumber
\end{eqnarray}
where $ \gamma_- = (\gamma_1^{\rm d}-\gamma_2^{\rm a})/4 $, $
\beta = \sqrt{ \epsilon^2 - \alpha^2} $, $ \alpha = \sqrt{
\kappa^2 + \gamma_+^2} $, and $ \gamma_+ = (\gamma_1^{\rm
d}+\gamma_2^{\rm a})/4 $. According to Eq.~(\ref{48}), all the
imaginary parts of four eigenfrequencies are equal. When damping
in mode 1 is stronger (weaker) than amplification in mode 2 [$
\gamma_1^{\rm d} > \gamma_2^{\rm a} $ ($ \gamma_1^{\rm d} <
\gamma_2^{\rm a} $)], the overall system is damped (amplified).

We note that the constant $ g $ is assumed real without the loss
of generality: It corresponds to a suitable choice of the phases
of the field operators $ \hat{a}_1 $ and $ \hat{a}_2 $ in
Eq.~(\ref{44}). On the other hand, the phases of possibly complex
constants $ \epsilon $ and $ \kappa $ have a good physical meaning
and influence the system dynamics to a certain extent. The
derivation of eigenfrequencies $ {\bf \Omega} $ in this most
general case results in the formulas that are derived from those
in~Eq.~(\ref{48}) by the formal replacement $ |\epsilon|
\rightarrow \epsilon $ and $ |\kappa| \rightarrow \kappa $. This
means that the spectrum of eigenfrequencies with its QEPs, QDPs,
and QHPs discussed below, remains the same. However, the formulas
for eigenvectors in Eq.~(\ref{49}), given below, have to be
replaced by more general ones in this case. We also note that no
QEP can be observed when different frequencies of the modes are
considered.

The unnormalized eigenvectors $ {\bf Y}_1 $, $ \bar{\bf Y}_1 $, $ {\bf Y}_2 $, and $ \bar{\bf Y}_2 $ belonging in turn
to the eigenfrequencies written in Eq.~(\ref{48}) are derived as follows
(assuming $ g \ge 0 $):
\begin{eqnarray}  
 {\bf Y}_{1,2}^T( \Omega_{1,2}^{\rm r})  &=& \Big\{-\epsilon\kappa +i\gamma_+(g \mp \alpha) \mp \alpha \Omega_{1,2}^{\rm r},
   \alpha^2  \mp g\alpha  \nonumber \\
 & & -i\gamma_+ \Omega_{1,2}^{\rm r} ,
  \pm\epsilon\alpha + \kappa \Omega_{1,2}^{\rm r},
  \kappa (g\mp\alpha) -i \gamma_+ \epsilon \Big\} , \nonumber \\
 \bar{\bf Y}_{1,2}( \Omega_{1,2}^{\rm r}) &=& {\bf Y}_{1,2}( -\Omega_{1,2}^{\rm r}) .
\label{49}
\end{eqnarray}

\begin{figure}   
  \centering
   \includegraphics[width=0.7\hsize]{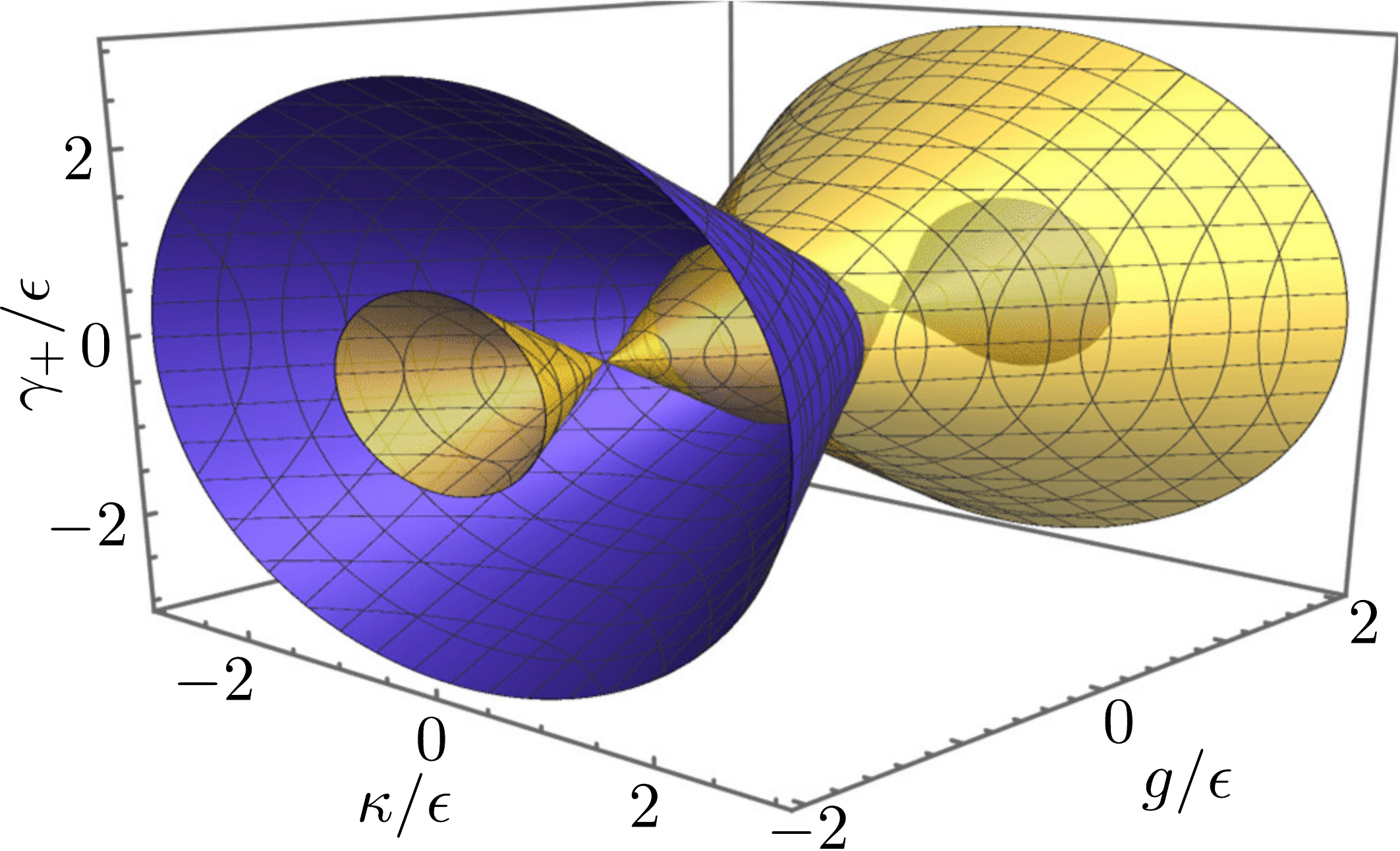}
  \caption{Two doubled cones with the shared axis defined by $ \kappa/\epsilon = \gamma_+/\epsilon = 0 $ identify the
    positions of QEPs in the space spanned by the parameters
    $ \gamma_+/\epsilon $, $ \kappa/\epsilon $, and $ g/\epsilon $. At $ g/\epsilon = 0 $,
    the two cones intersect which gives rise to the QHPs whose positions
    form a circle. We note that the surfaces of QEPs are symmetric with respect to the plane $ g/\epsilon = 0 $.}
\label{fig1}
\end{figure}

Assuming $ g \ne 0 $ and provided that the condition
\begin{equation}  
 \epsilon^2 - \left(\sqrt{\kappa^2+\gamma_+^2 } - g\right)^2
  = 0,
\label{50}
\end{equation}
or the condition
\begin{equation}  
 \epsilon^2 - \left(\sqrt{\kappa^2+\gamma_+^2 } + g\right)^2
  = 0,
\label{51}
\end{equation}
for the system parameters is fulfilled, two of the
eigenfrequencies coincide. This identifies a QEP for which the
eigenvectors corresponding to both eigenfrequencies coalesce. Each
of the above conditions forms a hypersurface of dimension 4 in the
space of independent parameters $ (\gamma_1^{\rm d},\gamma_2^{\rm
a},\epsilon,\kappa,g) $. Replacing the parameters $ \gamma_1^{\rm
d} $ and $ \gamma_2^{\rm a} $ by $ \gamma_+ $ and considering
linearity of the dynamics equations, the positions of QEPs form
two doubled concentric cones in the 3-dimensional space $ (
\gamma_+/\epsilon,\kappa/\epsilon,g/\epsilon) $ plotted in
Fig.~\ref{fig1}.

If $ g=0 $, two-fold degeneracy in eigenfrequencies occurs.
Real nonzero eigenfrequencies $ \Omega_{1,2}^{\rm r} $ exist only for $ \epsilon^2
- \kappa^2 - \gamma_+^2 > 0 $. When
\begin{equation}   
  \epsilon^2 - \kappa^2 - \gamma_+^2 = 0,
\label{52}
\end{equation}
all the four eigenfrequencies coincide. They form doubly
degenerate QEPs localized at a hypersurface of dimension 3 in the
parameter space $ (\gamma_1^{\rm d},\gamma_2^{\rm
a},\epsilon,\kappa) $ defined by the condition in Eq.~(\ref{52}).
The positions of the QEPs fulfilling Eq.~(\ref{52}) form the
circle with radius 1 in the space of parameters $
\gamma_+/\epsilon $ and $ \kappa/\epsilon $ shown in
Fig.~\ref{fig1} (the intersection of the yellow and blue cones).
At these QEPs, there exist two different eigenvectors each arising
from the original collapsing two dimensional spaces
\cite{PerinaJr2019c}. We have the QHPs in this case. We note that
the nonclassical properties of optical fields generated at and
around these QHPs were analyzed in \cite{PerinaJr2019c} where even
a more general Hamiltonian involving the additional Kerr nonlinear
terms was considered.

Now let us have a deeper look at the eigenfrequencies and their
spectral bifurcations that identify QEPs. To correctly identify
QEPs, we also need to know the eigenvectors that correspond to the
analyzed eigenfrequencies. The eigenvectors $ Y_1 $, $ \bar{Y}_1
$, $ Y_2 $, and $ \bar{Y}_2 $ given in Eq.~(\ref{49}), arising in
the diagonalization of the dynamics matrix of the
Heisenberg-Langevin equations (\ref{45}), and belonging in turn to
the eigenfrequencies $ \Omega_1 $, $ -\Omega_1^* $, $ \Omega_2 $,
and $ -\Omega_2^* $, may be used to form the eigenvectors of the
dynamics matrices of FOMs with increasing order. They directly
represent the eigenvectors of the first-order FOMs dynamics matrix
and, when formed into the supervector $ {\cal Y}^T \equiv ({\bf
Y}_1, \bar{\bf Y}_1, {\bf Y}_2, \bar{\bf Y}_2) $, they allow to
express the eigenvectors of the dynamics matrix of a $ p $th-order
FOMs via the tensor product $ \underbrace{ {\cal Y} \otimes \ldots
\otimes {\cal Y} }_{p} $ \cite{Arkhipov2022}. This allows us,
among other properties, to identify the number of eigenfrequencies
for a given order of FOMs and their degeneracies occurring at
QEPs.

The spectra of the eigenfrequencies and the numbers of
eigenvectors differ in the above-discussed two cases ($ g \ne 0 $
and $ g=0 $). Whereas three independent eigenvectors of the
dynamics matrix of the Heisenberg-Langevin equations occur at the
QEPs for $ g \ne 0 $, only two of the eigenvectors are found at
the QHPs when $ g=0 $. We note that, in both cases, all the
eigenvalues contribute to the dynamics of the original field
operators $ \hat{a}_1 $, $ \hat{a}_1^\dagger $, $ \hat{a}_2 $, and
$ \hat{a}_2^\dagger $ and so the analysis of the evolution of any
of them allows, in principle, to identify QEPs. In the following
eigenfrequency analysis, we pay a detailed attention to the
eigenfrequencies belonging to the FOMs up to the fourth order and
draw some conclusions concerning general orders.

\subsection{Spectra of eigenfrequencies for a single nondegenerate QEP}

We first consider the case for $ g \ne 0 $, in which a single
QEP with a double QEP degeneration occurs in the spectrum
of the dynamics matrix of the Heisenberg-Langevin equations.
At this QEP, three independent eigenvectors suffice in describing
the system evolution. In general, the eigenfrequency analysis of
the equations for the first-order FOMs provides four eigenfrequencies $ \pm \Omega_{1,2}^{\rm r}
-i\Omega^{\rm i} $. Two of them (say $ \pm \Omega_{1}^{\rm r}
-i\Omega^{\rm i} $) reveal the position of a QEP identified by the
condition $ \Omega_{1}^{\rm r} = 0 $.

This position of the QEP is also
indicated by the eigenfrequencies originating in the
analysis of higher-order FOMs.
We have in turn 4, 16, 64, and 256 moments of the first, second, third, and fourth orders.
However, as discussed above, some of these moments differ just in the positions of field operators. This means
that they are either equal or differ by the moments of lower orders if the involved field operators do not commute.
Taking into account this moment degeneracy, we may expect at maximum 4, 10, 20, and 35
different eigenfrequencies from the analysis of moments of the first, second, third, and fourth order.
Degeneracy of the moments is given in Tables~\ref{tab1} and \ref{tab2} for this case. This degeneracy is either mapped into
the multiplicity of the corresponding QEPs (forming QHPs from QEPs and being characterized by a QDP degeneracy) or results in higher
QEP degeneracies of QEPs (higher-order QEPs \cite{Arkhipov2021}).
Whereas the moments $ \langle \hat{b}_1 \hat{b}_2 \rangle $, $ \langle \hat{b}_2 \hat{b}_1 \rangle $,
$ \langle \hat{b}_1^\dagger \hat{b}_2 \rangle $, and $ \langle \hat{b}_2 \hat{b}_1^\dagger \rangle $ serve
as examples in the former case, the moments $ \langle \hat{b}_1^\dagger \hat{b}_1 \rangle $ and
$ \langle \hat{b}_1 \hat{b}_1^\dagger \rangle $ participate in forming a four-fold degenerated QEP (see Table~\ref{tab1}).
The eigenfrequencies attained by the analysis of the first-, second-, third- and fourth-order
FOMs are summarized in Tables~\ref{tab1} and \ref{tab2} depending on their ability to form QEPs.
We note that some of the eigenfrequencies summarized in Table~\ref{tab2}, that do not form QEPs, are degenerated, i.e.
they exhibit QDP degeneracy.

Provided that $ \Omega^{\rm i} \ne 0 $, we observe in the spectra
of eigenfrequencies in turn 1, 3, 6, and 10 bifurcations coming
from the behavior of the first-, second-, third-, and fourth-order
FOMs, as listed in Table~\ref{tab1}. Following Table~\ref{tab1},
there occur QHPs with QDP degeneracies 2, 6, and 12 considering in
turn the moments of the second, third, and fourth order. On the
other hand, the maximal QEP degeneracy of QEPs reached for the
first-, second-, \linebreak third-, and fourth-order FOMs equals
2, 4, 8 and 16. Here, we conclude that, in general, the analysis
of $ p $th-order FOMs gives QEPs with a $ 2^p $-fold degeneracy.
We note that some eigenfrequencies remain single as seen in
Table~\ref{tab2}. In Tables~\ref{tab1} and \ref{tab2}, we mention,
side by side with the eigenfrequencies, the corresponding moments
of the `diagonalized' field operators, as there is one-to-one
mapping between these moments and the structure of the
eigenfrequencies.

We note that, at the QEPs observed in the dynamics matrix of the
Heisenberg-Langevin equations, the number of independent
eigenvectors, given in Eq.~(\ref{49}), decreases from 4 to 3. This
is so as two eigenvectors have to coalesce at a nondegenerate QEP.
This results in the reduction of the complexity of the system
dynamics and leads to all physical effects discussed in relation
to QEPs. Considering higher-order FOMs, the number of independent
eigenvectors arising from the dynamics matrix of $ p $th-order
FOMs decreases from $ 4^p $ to $ 3^p $, which also gives the
maximal number of possibly different eigenfrequencies. Thus, the
dynamics of higher-order field-operator correlation functions is
more simplified at QEPs than that of the field mean operator
amplitudes.

The eigenfrequencies that are related to the moments containing
the `building block' $ \hat{b}_1^\dagger \hat{b}_1 $ (e.g., the
moments $ \langle \hat{b}_1^\dagger \hat{b}_1 \rangle $ and $
\langle \hat{b}_1^\dagger \hat{b}_1 \hat{b}_2\rangle $ in
Table~\ref{tab1}) form \emph{hidden QEPs.} The contribution to the
overall eigenfrequency of a higher-order FOM from this `building
block' equals zero as $ \Omega_1^{\rm r} - \Omega_1^{\rm r} = 0 $
independently of whether there is a QEP or not. However, the
presence of a QEP is identified by the reduction of the number of
eigenvectors by one that happens in this case without any
manifestation in the eigenfrequencies spectrum. Thus, no spectral
bifurcation commonly used for identifying QEPs is observed. We
note that, for the QEPs listed in Table~\ref{tab1}, such
eigenfrequencies form QEPs more than doubly degenerated, together
with other eigenfrequencies, and so these QEPs are still
identified in the spectrum by bifurcations.

In the usually discussed $ \cal PT $-symmetric systems, gain and loss
are in balance giving $ \Omega^{\rm i} = 0 $. In this case, multiplicties (i.e., QDP degeneracies) of QEPs may even be higher as
the imaginary parts of eigenfrequencies coming from different orders of FOMs coincide.
For example, the QEPs positioned at $ \pm \Omega_{1}^{\rm r} $ arise in the eigenfrequencies
of both the first- and third-order FOMs, in the latter case even with QDP degeneracy 6 (see Table~\ref{tab1}).
Such QEPs are called genuine QEPs, as discussed below.

We note that, when describing the evolution of FOMs of a given
order, we may neglect the redundant moments and keep just the
differential equations for the remaining ones. The
eigenfrequencies, originating in the analysis of such equations,
remain the same as those analyzed above and summarized in
Tables~\ref{tab1} and \ref{tab2}, but their multiplicities are
just 1. Also, the reduction in the number of involved moments
results in the change of the structure of the space of
eigenvectors. This reduction conceals the QHPs identified in the
last column of Table~\ref{tab1}. We may call such QHPs as the
\emph{induced QHPs} as they originate in the extended space of
FOMs that includes also the redundant moments. However, some QHPs,
which we refer to as \emph{the genuine QHPs}, still remain. These
occur for $ \Omega^{\rm i} = 0 $ and are formed by identical
eigenfrequencies with different eigenvectors arising in the
analysis of dynamics matrices for different orders FOMs (e.g., $
\langle \hat{b}_1 \rangle $ and $ \langle \hat{b}_1^\dagger
\rangle $ versus $ \langle \hat{b}_1 \hat{b}_2 \hat{b}_2^\dagger
\rangle $ and $ \langle \hat{b}_1^\dagger \hat{b}_2^\dagger
\hat{b}_2 \rangle $).

We also note that when reducing the number of necessary FOMs of
given order, we face the problem of non-commuting operators.
However, the FOMs that contain non-commuting operators at
different positions mutually differ by FOMs of orders lower by $
2,4,6, \ldots $: Each application of nontrivial commutation
relation reduced the moments order by two. When writing the
differential equations for the set of FOMs of given order, we
arrive at the equations similar to those found in Eqs.~(\ref{36}),
(\ref{39}) and (\ref{41}) that, however, contain additional terms
formed by lower-order FOMs at their r.h.s. Nevertheless, these
terms modify the nonhomogeneous solution of the equations
qualitatively in the same way as those arising in the fluctuating
Langevin forces, i.e. the solution is enriched by the terms
oscillating at the eigenfrequencies appropriate to these
lower-order FOMs. Thus the terms arising from the non-commuting
operators do not change the eigenfrequencies and we may apply the
above results concerning the eigenfrequencies also in this case.
On the other hand, the eigenvectors in both approaches naturally
differ. Their mutual relations \cite{Hu2021} were discussed in
relation to the quantum versus classical descriptions in
\cite{Arkhipov2022}.

\begin{table}    
\begin{center}
\begin{tabular}{|c|c|c|c|}
 \hline
  $ {\bf \Omega^{\rm i}_{\rm gen}} $ & $ {\bf \Omega^{\rm r}_{\rm gen}} $ & Moments & QDP deg.\\
 \hline
 \hline
  $ \Omega^{\rm i} $ & $ \Omega_2^{\rm r} $ & $ \langle \hat{b}_2 \rangle $ & 1 \\
 \cline{2-4}
  & $ -\Omega_2^{\rm r} $ & $ \langle \hat{b}_2^\dagger \rangle $ & 1 \\
 \hline
  $ 2\Omega^{\rm i} $ & 0 &  $ \langle \hat{b}_2^\dagger \hat{b}_2 \rangle $ & 2 \\
 \cline{2-4}
   & $ 2\Omega_2^{\rm r} $ & $ \langle \hat{b}_2^2 \rangle $ & 1 \\
 \cline{2-4}
   & $ - 2\Omega_2^{\rm r} $ & $ \langle \hat{b}_2^{\dagger 2} \rangle $ & 1 \\
 \hline
  $ 3\Omega^{\rm i} $ &  $ \Omega_2^{\rm r} $ & $ \langle \hat{b}_2^{\dagger 2} \hat{b}_2 \rangle $ & 3\\
 \cline{2-4}
   & $ -\Omega_2^{\rm r} $  & $ \langle \hat{b}_2^\dagger \hat{b}_2^2 \rangle $ & 3\\
 \cline{2-4}
   & $ 3\Omega_2^{\rm r} $ & $ \langle \hat{b}_2^3 \rangle $ & 1 \\
 \cline{2-4}
   & $ -3\Omega_2^{\rm r} $ & $ \langle \hat{b}_2^{\dagger 3} \rangle $ & 1 \\
\hline
  $ 4\Omega^{\rm i} $ & 0 & $ \langle \hat{b}_2^{\dagger 2} \hat{b}_2^2 \rangle $ & 6 \\
 \cline{2-4}
  & $ 2\Omega_2^{\rm r} $  & $ \langle  \hat{b}_2^\dagger \hat{b}_2^3\rangle $ &  4\\
 \cline{2-4}
  & $ -2\Omega_2^{\rm r} $ & $ \langle \hat{b}_2^{\dagger 3} \hat{b}_2 \rangle $ & 4\\
 \cline{2-4}
   & $ 4\Omega_2^{\rm r} $ & $ \langle \hat{b}_2^4 \rangle $ & 1 \\
 \cline{2-4}
   & $ -4\Omega_2^{\rm r} $ & $ \langle \hat{b}_2^{\dagger 4} \rangle $ & 1 \\
 \hline
\end{tabular}
\end{center}
\caption{Real and imaginary parts of the complex eigenfrequencies
$ \Omega_{\rm gen}^{\rm r} - i
  \Omega_{\rm gen}^{\rm i} $ derived from the equations for the
  FOMs up to fourth order which \emph{do not
  indicate} a QEP for $ g \ne 0 $. The corresponding moments in the
  `diagonalized' field operators are written together with their QDP degeneracies resulting from
  different orderings of field operators.}
\label{tab2}
\end{table}


\subsection{Spectra of eigenfrequencies for a doubly degenerated QEP - QHP}

The squeezing-effect  part of the Hamiltonian $ \hat{H}^0 $ in Eq.~(\ref{44})
is often not considered ($ g = 0 $). This leads to two doubly
degenerated eigenfrequencies $ \pm \Omega^{\rm r} -i\Omega^{\rm i}
$, when the dynamics of the first-order FOMs is investigated. They form a QHP occurring
directly in the dynamics matrix of the Heisenberg-Langevin equations. This QDP degeneracy is then directly transformed
into the diabolical degeneracies of eigenfrequencies arising from the analysis of higher-order FOMs. We may call such QHP the \emph{inherited
QHP.} This inherited QDP degeneracy considerably reduces the
number of different spectral eigenfrequencies provided by higher-order FOMs, at the expense of their increasing degeneracies.
We note that this behavior was observed in \cite{Minganti2019},
where the spectrum of the Liouvillian of a simplified two-mode bosonic
system with $ g=\kappa=0 $ was numerically analyzed. The obtained
eigenfrequencies, their QDP and QEP degeneracies
and the corresponding QHPs
are summarized in Table~\ref{tab3}, together with the corresponding `diagonalized'
FOMs.
\begin{table}[t]    
\begin{center}
\begin{tabular}{|c|c|c|c|c|c|}
 \hline
  $ {\bf \Omega^{\rm i}_{\rm gen}} $ & $ {\bf \Omega^{\rm r}_{\rm gen}} $ & Moments & Moment & Partial & Partial\\
   &  &  & deg. & QDP x & QDP x \\
   &  &  &  & QEP deg. & QEP deg.\\
 \hline
 \hline
  $ \Omega^{\rm i} $ & $ \pm \Omega^{\rm r} $ & $ \langle \hat{b}_1 \rangle $, $ \langle \hat{b}_1^\dagger \rangle $ & 1 & 1x2 & 2x2\\
 \cline{3-5}
  &  & $ \langle \hat{b}_2 \rangle $, $ \langle \hat{b}_2^\dagger \rangle $ & 1 & 1x2 & \\
 \hline
  $ 2\Omega^{\rm i} $ & $ \pm 2\Omega^{\rm r} $ & $ \langle \hat{b}_1 \hat{b}_2 \rangle $, $ \langle \hat{b}_1^\dagger \hat{b}_2^\dagger \rangle $ & 2 & 2x4 & 4x4 \\
 \cline{2-4}
      & $ \Omega^{\rm r} - \Omega^{\rm r} $  &  $ \langle \hat{b}_1^\dagger \hat{b}_2 \rangle $ & 2 & & \\
 \cline{2-4}
      & $ \Omega^{\rm r} - \Omega^{\rm r} $  &  $ \langle \hat{b}_1 \hat{b}_2^\dagger \rangle $ & 2 &  & \\
 \cline{2-5}
  & $ \pm 2\Omega^{\rm r} $ & $ \langle \hat{b}_1^2 \rangle $, $ \langle \hat{b}_1^{\dagger 2}\rangle $ & 1 & 1x4 & \\
 \cline{2-4}
  & $ \Omega^{\rm r} - \Omega^{\rm r} $  & $ \langle \hat{b}_1^\dagger \hat{b}_1 \rangle $ & 2 & & \\
 \cline{2-5}
  & $ \pm 2\Omega^{\rm r} $ & $ \langle \hat{b}_2^2 \rangle $, $ \langle \hat{b}_2^{\dagger 2}\rangle $ & 1 & 1x4 & \\
 \cline{2-4}
      & $ \Omega^{\rm r} - \Omega^{\rm r} $  &  $ \langle \hat{b}_2^\dagger \hat{b}_2 \rangle $ & 2 & & \\
 \hline
  $ 3\Omega^{\rm i} $ & $ \pm 3 \Omega^{\rm r} $  & $ \langle \hat{b}_1^2 \hat{b}_2 \rangle $, $ \langle \hat{b}_1^{\dagger 2} \hat{b}_2^\dagger \rangle $ & 3 & 3x8 & 8x8 \\
 \cline{2-4}
  & $ \pm \Omega^{\rm r} $ & $ \langle \hat{b}_1^2 \hat{b}_2^\dagger \rangle $, $ \langle \hat{b}_1^{\dagger 2} \hat{b}_2 \rangle $ & 3 & & \\
 \cline{3-4}
  &  & $ \langle \hat{b}_1^\dagger \hat{b}_1 \hat{b}_2 \rangle $, $ \langle \hat{b}_1^\dagger \hat{b}_1 \hat{b}_2^\dagger \rangle $ & 6 & & \\
 \cline{2-5}
  & $ \pm 3 \Omega^{\rm r} $  & $ \langle \hat{b}_1 \hat{b}_2^2 \rangle $, $ \langle \hat{b}_1^\dagger \hat{b}_2^{\dagger 2} \rangle $ & 3 & 3x8 & \\
 \cline{2-4}
  & $ \pm \Omega^{\rm r} $ & $  \langle \hat{b}_1^\dagger \hat{b}_2^2 \rangle $, $ \langle \hat{b}_1 \hat{b}_2^{\dagger 2} \rangle $ & 3 & & \\
 \cline{3-4}
  & & $ \langle \hat{b}_1 \hat{b}_2^\dagger \hat{b}_2 \rangle $, $ \langle \hat{b}_1^\dagger \hat{b}_2^\dagger \hat{b}_2  \rangle $ & 6 & & \\
 \cline{2-5}
   & $ \pm 3 \Omega^{\rm r} $ & $ \langle \hat{b}_1^3 \rangle $, $ \langle \hat{b}_1^{\dagger 3} \rangle $ & 1 & 1x8 & \\
 \cline{2-4}
  & $ \pm \Omega^{\rm r} $ & $ \langle \hat{b}_1^\dagger  \hat{b}_1^2 \rangle $, $ \langle \hat{b}_1^{\dagger 2} \hat{b}_1 \rangle $ & 3 & & \\
 \cline{2-5}
  & $ \pm 3 \Omega^{\rm r} $  & $ \langle \hat{b}_2^3 \rangle $, $ \langle \hat{b}_2^{\dagger 3} \rangle $ & 1 & 1x8 & \\
 \cline{2-4}
  & $ \pm \Omega^{\rm r} $ & $ \langle \hat{b}_2^\dagger \hat{b}_2^2 \rangle $, $ \langle \hat{b}_2^{\dagger 2} \hat{b}_2 \rangle $ & 3 & & \\
 \cline{2-5}
\hline
  $ 4\Omega^{\rm i} $ & $ \pm 4 \Omega^{\rm r} $ & $ \langle \hat{b}_1^3 \hat{b}_2 \rangle $, $ \langle \hat{b}_1^{\dagger 3} \hat{b}_2^\dagger \rangle $ & 4 & 4x16 & 16x16\\
 \cline{2-4}
  & $ \pm 2 \Omega^{\rm r} $ & $ \langle \hat{b}_1^3 \hat{b}_2^\dagger \rangle $, $ \langle \hat{b}_1^{\dagger 3} \hat{b}_2 \rangle $ & 4 & & \\
 \cline{3-4}
  &  & $ \langle \hat{b}_1^\dagger \hat{b}_1^2  \hat{b}_2  \rangle $, $ \langle \hat{b}_1 \hat{b}_1^{\dagger 2} \hat{b}_2^\dagger \rangle $ & 12 & & \\
 \cline{2-4}
      & $ 2 \Omega^{\rm r} - 2 \Omega^{\rm r} $ &  $ \langle \hat{b}_1^\dagger \hat{b}_1^2 \hat{b}_2^\dagger \rangle $ , $ \langle \hat{b}_1^{\dagger 2} \hat{b}_1  \hat{b}_2 \rangle $ & 12 & & \\
 \cline{2-5}
 & $ \pm 4 \Omega^{\rm r} $ & $ \langle \hat{b}_1 \hat{b}_2^3 \rangle $, $ \langle \hat{b}_1^\dagger \hat{b}_2^{\dagger 3} \rangle $ & 4 & 4x16 & \\
 \cline{2-4}
  & $ \pm 2 \Omega^{\rm r} $ & $ \langle \hat{b}_1^\dagger \hat{b}_2^3  \rangle $, $ \langle \hat{b}_1 \hat{b}_2^{\dagger 3} \rangle $ & 4 & & \\
 \cline{3-4}
  & & $ \langle \hat{b}_1 \hat{b}_2^\dagger \hat{b}_2^2  \rangle $, $ \langle \hat{b}_1^\dagger \hat{b}_2^{\dagger 2} \hat{b}_2 \rangle $ & 12 & & \\
 \cline{2-4}
      & $ 2 \Omega^{\rm r} - 2 \Omega^{\rm r} $  &  $ \langle \hat{b}_1 \hat{b}_2^{\dagger 2} \hat{b}_2 \rangle $ ,  $ \langle \hat{b}_1^\dagger \hat{b}_2^2 \hat{b}_2^\dagger \rangle $ & 12 & & \\
 \cline{2-5}
  & $ \pm 4 \Omega^{\rm r} $ & $ \langle \hat{b}_1^2 \hat{b}_2^2 \rangle $, $ \langle \hat{b}_1^{\dagger 2} \hat{b}_2^{\dagger 2} \rangle $ & 6 & 6x16 & \\
 \cline{2-4}
   & $ 2 \Omega^{\rm r} - 2 \Omega^{\rm r} $  &  $ \langle \hat{b}_1^2 \hat{b}_2^{\dagger 2} \rangle $,  $ \langle \hat{b}_1^{\dagger 2} \hat{b}_2^2 \rangle $ & 6 & & \\
 \cline{2-4}
  & $ \pm 2 \Omega^{\rm r} $ & $ \langle \hat{b}_1^2 \hat{b}_2^\dagger \hat{b}_2  \rangle $, $ \langle \hat{b}_1^{\dagger 2} \hat{b}_2^\dagger \hat{b}_2 \rangle $ & 12 & & \\
 \cline{3-4}
  &  & $ \langle \hat{b}_1^\dagger \hat{b}_1  \hat{b}_2^2  \rangle $, $ \langle \hat{b}_1^\dagger \hat{b}_1 \hat{b}_2^{\dagger 2} \rangle $ & 12 & & \\
 \cline{2-4}
     & $ 2 \Omega^{\rm r} - 2 \Omega^{\rm r} $  &  $ \langle \hat{b}_1^\dagger \hat{b}_1 \hat{b}_2^\dagger \hat{b}_2 \rangle $ & 24 & & \\
 \cline{2-5}
  & $ \pm 4 \Omega^{\rm r} $ & $ \langle \hat{b}_1^4 \rangle $, $ \langle \hat{b}_1^{\dagger 4} \rangle $ & 1 & 1x16 & \\
 \cline{2-4}
   & $ \pm 2 \Omega^{\rm r} $ & $ \langle \hat{b}_1^\dagger \hat{b}_1^3 \rangle $, $ \langle \hat{b}_1^{\dagger 3} \hat{b}_1 \rangle $ & 4 & & \\
 \cline{2-4}
  & $ 2 \Omega^{\rm r} - 2 \Omega^{\rm r} $  & $ \langle \hat{b}_1^{\dagger 2} \hat{b}_1^2 \rangle $ & 6 & & \\
 \cline{2-5}
  & $ \pm 4 \Omega^{\rm r} $ & $ \langle \hat{b}_2^4 \rangle $, $ \langle \hat{b}_2^{\dagger 4} \rangle $ & 1 & 1x16 & \\
 \cline{2-4}
  & $ \pm 2 \Omega^{\rm r} $ & $ \langle \hat{b}_2^\dagger \hat{b}_2^3 \rangle $, $ \langle \hat{b}_2^{\dagger 3} \hat{b}_2 \rangle $ & 4 & & \\
 \cline{2-4}
      & $ 2 \Omega^{\rm r} - 2 \Omega^{\rm r} $ & $ \langle \hat{b}_2^{\dagger 2} \hat{b}_2^2 \rangle $ & 6 & & \\
  \hline
\end{tabular} 
\end{center}
\vspace{-5mm}
 \caption{Real and imaginary parts of the complex eigenfrequencies $
\Omega_{\rm gen}^{\rm r} - i
  \Omega_{\rm gen}^{\rm i} $ derived from the equations for the
  FOMs up to fourth order which
  \emph{indicate} a QHP for $ g = 0 $. The corresponding moments in the
  `diagonalized' field operators are written together with their degeneracies coming from
  different orderings of field operators. QDP degeneracies of QHPs (partial QDP degeneracies) derived from
  the indicated FOMs and QEP degeneracies of the constituting QEPs are given.}
\label{tab3}
\end{table}
According to Table~\ref{tab3}, the QEPs found for the dynamics
matrix of $ p $th-order FOMs exhibit a $ 2^p $-fold QEP
degeneracy. Due to the inherited spectral degeneracy, the
frequencies of all QEPs belonging to $ p $th-order FOMs equal zero
($\Omega^{\rm r} = 0 $), which results in the occurrence of the
QHP with a QDP degeneracy $ 2^p $ formed by $ 2^p $ QEPs with a $
2^p $-fold QEP degeneracy, as shown in Table~\ref{tab3}. This
results from the fact that the number of independent eigenvectors
of the dynamics matrix of the Heisenberg-Langevin equations
decreases from 4 to 2 at QHPs.

As seen in Table~\ref{tab3}, we have the hidden QEPs/QHPs also in
this case. They occur not only in relation to the moments of a
single mode (e.g., $ \langle \hat{b}_1^\dagger \hat{b}_1 \rangle $
and $ \langle \hat{b}_1 \hat{b}_1^\dagger \rangle $), but also
when cross moments of different modes are considered (e.g., $
\langle \hat{b}_1^\dagger \hat{b}_2 \rangle $ and $ \langle
\hat{b}_1 \hat{b}_2^\dagger \rangle $, or $ \langle \hat{b}_2
\hat{b}_1^\dagger \rangle $ and $ \langle \hat{b}_2^\dagger
\hat{b}_1 \rangle $). Apart from the hidden QEPs, we observe
spectral bifurcations at $ \pm p \Omega^{\rm r} $, $ \pm (p-2)
\Omega^{\rm r} $, $ \pm (p-4) \Omega^{\rm r} $, $ \ldots $ from
the $ p $th-order FOMs.

Provided that we exclude the redundant FOMs from the description, the QDP degeneracy of QHPs
identified in Table~\ref{tab3}, in general, lowers. Whereas it remains 2 for the first-order FOMs,
it decreases from $ 2^p $ to $ p+1 $ for $ p $th-order FOMs.

\hspace{2mm}

Finally, we discuss some properties of eigenfrequencies when the
FOMs of all orders are considered. Provided that the gain and loss
are in balance, we have $ {\Omega}^{\rm i} = 0 $ and we can
directly compare the eigenfrequencies arising in the dynamics
matrices of FOMs of different orders. Then the QEPs are localized
with the help of pairs of eigenfrequencies $ \pm p \Omega_1^{\rm
r} $ ($ g \ne 0 $) and $ \pm p \Omega^{\rm r} $ ($ g = 0 $) for $
p=1,2,\ldots $ occurring infinitely-many times in the spectrum.
For $ g \ne 0 $, the other pairs of eigenfrequencies, as
explicitly written in Table~\ref{tab1}, are, among others, also
found in the spectrum with infinite degeneration.

At the end, we note that, when the eigenfrequency analysis is
accompanied by the determination of the corresponding
eigenvectors, we can discuss the modes behavior at a general level
using the quantities based on the determined FOMs. For example,
phase squeezing is revealed by the behavior of the second-order
FOMs \cite{Luks1988,Lvovsky2009}, whereas sub-Poissonian
photon-number statistics \cite{Mandel1995} of the modes and their
sub-shot-noise photon-number correlations \cite{Bondani2007} are
quantified by the fourth-order FOMs. Also different types of
nonclassicalities can be discussed \cite{PerinaJr2022}.

We also note that our approach relies on the linear
Heisenberg-Langevin equations. Nevertheless, it may be
successfully applied also in investigations of quantum systems
described by the nonlinear Heisenberg-Langevin equations provided
that a suitable operator linearization of the nonlinear operator
equations is applied, e.g. around a stationary state or a
classical time-dependent solution
\cite{PerinaJr2019c,PerinaJr2019b,PerinaJr2016b,PerinaJr2000}.

\clearpage

\section{Conclusions}

We have shown that the eigenfrequency analysis of the Liouvillians of
open quantum systems can alternatively be performed in the space
of operators of measurable quantities provided that they form a
complete basis. This is especially important for systems defined in
infinite-dimensional Liouville spaces including those formed by
the interacting bosonic fields. Considering a damped two-level atom,
we have demonstrated the equivalence of both approaches in
obtaining the system eigenfrequencies and positions of quantum exceptional points.
Analysing the dynamical equations of field-operator moments of
general $ M $-mode fields, we have revealed the structure of
eigenfrequencies attainable from dynamical equations for a given
order of the field-operator moments. This shed light to the
occurrence of quantum exceptional points identified from the obtained eigenfrequencies:
All quantum exceptional points are recognized already from the eigenfrequencies obtained
from the first-order field-operator moments. The
eigenfrequencies obtained from higher-order field-operator
moments are important in revealing multiple (i.e., diabolical) degeneracies of these quantum exceptional points only.
We have developed a general approach to analyze a two-mode
bosonic system described by a general quadratic Hamiltonian. In
its general configuration, two distinct sets of quantum exceptional points occur for nonzero mode squeezing that,
however, collapse into a single set with quantum hybrid diabolical exceptional points, when mode
squeezing is not considered. In the analysis, we have observed the inherited, genuine and induced
quantum hybrid diabolical exceptional points. Moreover, the hidden quantum exceptional points, whose
presence is not directly inferred from the behavior of eigenfrequency spectra, were identified.

The consideration of the Heisenberg-Langevin equations for the
operators of measurable quantities and the derived dynamical
equations for field-operator moments represent a convenient
starting point for the system eigenfrequency analysis that allows
to reveal the eigenfrequencies of open quantum systems encoded in
their Liouvillians. We believe that this approach is qualitatively less demanding
compared to a direct diagonalization of the Liouvillians, at least
when the linear Heisenberg-Langevin equations describe the analyzed system. This
approach paves the way to a general and detailed analysis of quantum exceptional points in open
quantum infinitely-dimensional systems.

\acknowledgments We thank Ievgen I. Arkhipov for fruitful
discussions on EPs. J.P. thanks Anton\'\i n Luk\v{s} for
discussing the formalism and careful reading the manuscript. J.P.
acknowledges the support from M\v{S}MT \v{C}R projects Nos.
CZ.02.2.69/0.0/0.0/18\_053/0016919 and CZ.02.1.01/0.0/
0.0/16\_019/0000754. A.M., A.K.-K., and G.C. were supported by the
Polish National Science Centre (NCN) under the Maestro Grant No.
DEC-2019/34/A/ST2/00081.

\appendix

\section{Correspondence between the master equation
and the Heisenberg-Langevin equations}

We consider a harmonic oscillator with frequency $ \omega $ damped
by the interaction with a system of two-level atoms effectively
described by an `average' two-level atom with its rasing ($
\hat{\sigma}_+ $) and lowering operators ($ \hat{\sigma}_- $). The
second-order perturbation solution of the Liouville equation for
the overall statistical operator, when traced over the reservoir
stationary state, leads to the following master
equation for the reduced statistical operator $ \hat{\rho}^{\rm d}
$ of the damped oscillator \cite{Vogel2006,Meystre2007}:
\begin{equation}  
 \frac{ \partial \hat{\rho}^{\rm d} }{\partial t} = -i\omega [ \hat{a}^\dagger\hat{a},\hat{\rho}^{\rm d}]
  +  \frac{\gamma^{{\rm d}'}}{2} \left( [\hat{a},\hat{\rho}^{\rm
  d}\hat{a}^\dagger] -  [\hat{a}^\dagger,\hat{a}\hat{\rho}^{\rm
  d}] \right),
\label{A1}
\end{equation}
where $ \hat{a} $ ($ \hat{a}^\dagger $) stands for the oscillator
annihilation (creation) operator. We assume the `average'
reservoir two-level atom in the ground state, i.e. $ \langle
\hat{\sigma}_- \hat{\sigma}_+ \rangle_{\rm R} = 1 $ and $ \langle
\hat{\sigma}_+ \hat{\sigma}_- \rangle_{\rm R} = 0 $, and so we
have $ \gamma^{{\rm d}'} \equiv \gamma^{\rm d} \langle
\hat{\sigma}_- \hat{\sigma}_+ \rangle_{\rm R} = \gamma^{\rm d} $
using the damping constant $\gamma^{\rm d} $.

Using the Glauber-Sudarshan representation of statistical operator
\cite{Glauber1963,Sudarshan1963}
\begin{displaymath}
  \hat{\rho} = \int d^{2}\alpha \, \Phi_{\cal
 N}(\alpha,\alpha^*) |\alpha\rangle \langle\alpha|
\end{displaymath}
written in the basis of coherent states $
|\alpha\rangle $ and applying the identities $ [\hat{a},
f(\hat{a},\hat{a}^\dagger)] =
\partial f(\hat{a},\hat{a}^\dagger) / \partial \hat{a}^\dagger $,
$ [g(\hat{a},\hat{a}^\dagger),\hat{a}^\dagger] =
\partial g(\hat{a},\hat{a}^\dagger) / \partial \hat{a} $, we
transform the master equation (\ref{A1}) into the
corresponding Fokker-Planck equation \cite{Risken1996,Perina1991}:
\begin{equation}   
 \frac{ \partial \Phi_{\cal N}^{\rm d} }{\partial t} = \left(\frac{\gamma^{{\rm d}'}}{2} -i\omega\right)
  \frac{\partial (\alpha \Phi_{\cal N}^{\rm d}) }{\partial \alpha} +
  \left(\frac{\gamma^{{\rm d}'}}{2} + i\omega\right)
  \frac{\partial (\alpha^* \Phi_{\cal N}^{\rm d}) }{\partial \alpha^*}.
\label{A2}
\end{equation}
The Fokker-Planck equation (\ref{A2}) is then equivalent to the
set of the Heisenberg-Langevin equations,
\begin{eqnarray}   
 \frac{d\hat{a}(t)}{dt} &=& \left(-i\omega - \frac{\gamma^{{\rm d}'}}{2}\right) \hat{a}(t) +
 \hat{L}^{\rm d}(t), \nonumber \\
 \frac{d\hat{a}^\dagger(t)}{dt} &=& \left(i\omega - \frac{\gamma^{{\rm d}'}}{2}\right) \hat{a}^\dagger(t) +
 \hat{L}^{{\rm d}\dagger}(t),
\label{A3}
\end{eqnarray}
with the stochastic Langevin operator forces $ \hat{L}^{\rm d} $
and $ \hat{L}^{{\rm d}\dagger} $ endowed with the following
Gaussian and Markovian properties:
\begin{eqnarray}   
 \langle \hat{L}^{\rm d}(t) \rangle = \langle
  \hat{L}^{{\rm d}\dagger}(t) \rangle &=& 0 ,  \nonumber \\
 \langle \hat{L}^{\rm d}(t) \hat{L}^{{\rm d}\dagger}(t') \rangle &=&
  \gamma^{{\rm d}'} \delta(t-t') ,  \nonumber \\
 \langle \hat{L}^{{\rm d}\dagger}(t) \hat{L}^{\rm d}(t') \rangle &=&
  0 .
\label{A4}
\end{eqnarray}

On the other hand, when the harmonic oscillator interacts with the
`average' reservoir two-level atom in the excited state, i.e. $
\langle \hat{\sigma}_- \hat{\sigma}_+ \rangle_{\rm R} = 0 $ and $
\langle \hat{\sigma}_+ \hat{\sigma}_- \rangle_{\rm R} = 1 $, it is
amplified. Its master equation for the reduced
statistical operator $ \hat{\rho}^{\rm a} $ is derived in the
form:
\begin{equation}  
 \frac{ \partial \hat{\rho}^{\rm a} }{\partial t} = -i\omega [ \hat{a}^\dagger\hat{a},\hat{\rho}^{\rm a}]
  +  \frac{\gamma^{{\rm a}'}}{2} \left( [\hat{a}^\dagger,\hat{\rho}^{\rm
  a}\hat{a}] -  [\hat{a},\hat{a}^\dagger\hat{\rho}^{\rm
  a}] \right).
\label{A5}
\end{equation}
In Eq.~(\ref{A5}), $ \gamma^{{\rm a}'} \equiv \gamma^{\rm a}
\langle \hat{\sigma}_+ \hat{\sigma}_- \rangle_{\rm R} =
\gamma^{\rm a} $, where $ \gamma^{\rm a} $ denotes the
amplification constant. The corresponding Fokker-Planck equation,
\begin{eqnarray}   
 \frac{ \partial \Phi_{\cal N}^{\rm a} }{\partial t} = \left(-\frac{\gamma^{{\rm a}'}}{2} -i\omega\right)
  \frac{\partial (\alpha \Phi_{\cal N}^{\rm a}) }{\partial \alpha} + \left( -\frac{\gamma^{{\rm a}'}}{2} + i\omega\right) 
  \frac{\partial (\alpha^* \Phi_{\cal N}^{\rm a}) }{\partial \alpha^*}  + \gamma^{{\rm a}'}
  \frac{\partial^2 \Phi_{\cal N}^{\rm a} }{\partial \alpha \partial
  \alpha^* } ,
\label{A6}
\end{eqnarray}
is then equivalent to the set of the Heisenberg-Langevin equations,
\begin{eqnarray}   
 \frac{d\hat{a}(t)}{dt} &=& \left(-i\omega + \frac{\gamma^{{\rm a}'}}{2}\right) \hat{a}(t) +
 \hat{L}^{\rm a}(t), \nonumber \\
 \frac{d\hat{a}^\dagger(t)}{dt} &=& \left(i\omega + \frac{\gamma^{{\rm a}'}}{2}\right) \hat{a}^\dagger(t) +
 \hat{L}^{{\rm a}\dagger}(t),
\label{A7}
\end{eqnarray}
with the stochastic Langevin operator forces, $ \hat{L}^{\rm a} $
and $ \hat{L}^{{\rm a}\dagger} $, obeying the following
properties:
\begin{eqnarray}   
 \langle \hat{L}^{\rm a}(t) \rangle = \langle
  \hat{L}^{{\rm a}\dagger}(t) \rangle &=& 0 ,  \nonumber \\
 \langle \hat{L}^{\rm a}(t) \hat{L}^{{\rm a}\dagger}(t') \rangle &=&
  0 ,  \nonumber \\
 \langle \hat{L}^{{\rm a}\dagger}(t) \hat{L}^{\rm a}(t') \rangle &=&
  \gamma^{{\rm a}'} \delta(t-t') .
\label{A8}
\end{eqnarray}


\end{document}